\documentclass[twocolumn,tighten,twocolappendix]{aastex701}
\usepackage{graphicx}
\usepackage{amsmath}
\usepackage{natbib}
\usepackage{amssymb}
\usepackage{ulem}
\usepackage{sidecap}
\usepackage{hyperref}
\usepackage{physics}

\shorttitle{Image of TDE Radio Flare}
\shortauthors{Matsumoto}

\begin{document}

\title{VLBI Diagnostics of Off-axis Jets in Radio Flares of Tidal Disruption Events}

\author[0000-0002-9350-6793]{Tatsuya Matsumoto}
\affil{Department of Astronomy, School of Science, The University of Tokyo, Bunkyo-ku, Tokyo 113-0033, Japan}
\email[show]{matsumoto@astron.s.u-tokyo.ac.jp}

\begin{abstract}
The origin of late-time radio flares in tidal disruption events remains unclear. In particular, the peculiar radio flare observed in AT2018hyz has motivated two leading scenarios: a delayed outflow launched $\sim1000\,\rm days$ after discovery, or an off-axis relativistic jet directed far from our line of sight. Very long baseline interferometry (VLBI) imaging provides the most direct way to distinguish between these scenarios. In this paper, we calculate synthetic radio images for both models and examine their observational signatures. The motion of the emission centroid is the most powerful diagnostic for breaking the degeneracy. In the delayed-outflow scenario, the centroid motion is confined within a non-relativistic distance, whereas in the off-axis jet scenario it exhibits apparent superluminal motion. Detecting such superluminal motion would therefore provide a smoking-gun signature of the off-axis jet interpretation. We also find that the jet image exhibits characteristic features, including a non-monotonic evolution of the image aspect ratio. These results are expected to be generic and applicable to other jetted explosions, such as microquasars and gamma-ray bursts.
\end{abstract}


\section{Introduction}
Recent observations have revealed that a significant fraction of optical tidal disruption events \citep[TDEs;][]{Rees1988} are accompanied by radio flares appearing as late as $\sim1000\,\rm days$ after discovery (\citealt{Cendes+2024}; see also \citealt{Horesh+2021,Horesh+2021b,Cendes+2022b,Goodwin+2023b,Christy+2024,ZhangFabao+2024,Golay+2025,Hajela+2025,Goodwin+2025}). Several ideas have been proposed for the origin of these late-time radio flares, including delayed outflows produced by delayed disk formation or state transitions in the disk \citep{Cendes+2022b,Alexander+2026,WuSamantha+2026}, prompt outflows interacting with a circumnuclear medium (CNM) with a non-monotonic density profile \citep{Matsumoto&Piran2024,Zhuang+2025}, and off-axis relativistic jets \citep{Matsumoto&Piran2023,Sfaradi+2024}. Despite these efforts, their physical origin remains uncertain.

Among the late-time radio flares, AT2018hyz stands out as a particularly remarkable event \citep{Cendes+2022b,Sfaradi+2024,Cendes+2026}. After $\sim1000\,\rm days$ of non-detection, a radio flare suddenly emerged with a sharply rising light curve and an unusually high luminosity, comparable to that of jetted TDEs \citep{Eftekhari+2018,Cendes+2021}. As one of the earliest identified examples of late-time flares, this event was initially interpreted as being powered by a delayed outflow \citep{Cendes+2022b}. \cite{Matsumoto&Piran2023} further demonstrated that an off-axis relativistic jet \citep[e.g.,][]{Giannios&Metzger2011} can also reproduce the observed properties. At present, both scenarios remain viable and cannot be distinguished using the radio light curve and spectrum alone \citep{Cendes+2026}.

\cite{Matsumoto&Piran2023} proposed that radio imaging with very long baseline interferometry (VLBI) could provide a direct diagnostic of the origin of the radio emission in AT2018hyz. VLBI observations have played a crucial role in revealing the nature of relativistic transients, including Galactic microquasars (\citealt{Mirabel&Rodriguez1994,Tingay+1995}; see also \citealt{Fender2006}) and binary neutron star (BNS) mergers \citep{Mooley+2018b,Ghirlanda+2019}. In particular, the BNS merger GW170817 has heightened interest in radio imaging and motivated numerous theoretical studies, using both (semi-)analytical \citep{Gill&Granot2018b,Fernandez+2022,GovreenSegal&Nakar2023,Sadeh+2024} and numerical approaches \citep{Zrake+2018,Granot+2018,Fernandez+2022,Nedora+2023}. The same techniques are applicable to TDEs.

In this paper, we compute radio images of late-time radio flares in TDEs using a semi-analytical framework. In the context of TDEs, several previous studies have calculated synthetic images based on simulations \citep[e.g.,][]{Mimica+2015,Hu+2025,Mou&Shu2025}, but did not carry out a detailed analysis of their properties. Moreover, recent VLBI observations of TDE radio flares \citep{Golay+2025,Hajela+2025} further motivate a detailed theoretical investigation of their expected imaging signatures. Our approach allows us to isolate key physical effects and clarify the observational diagnostics that can distinguish between different scenarios.

The structure of the paper is as follows. In Section~\ref{sec:method}, we describe the method used to calculate the radio images as well as light curve and spectrum. In Section~\ref{sec:result}, we present radio images of AT2018hyz based on two different models. We discuss the very late-time evolution of the radio images and their implications in Section~\ref{sec:long time}, and summarize our findings in Section~\ref{sec:summary}.

\section{Method}
\label{sec:method}

Light curves and images of synchrotron-emitting outflows have been calculated by many authors at various levels of approximation. In this study, we assume that the outflow is piecewise spherical; that is, each fluid element evolves as though it were part of a spherically symmetric outflow. We further assume that the emitting region is a thin shell, and do not consider any angular structure \citep[but see e.g.,][]{Tchekhovskoy+2014,Teboul&Metzger2023,Lu+2024}.

\begin{figure}
    \centering
    \includegraphics[width=85mm,bb=0 0 1829 1353]{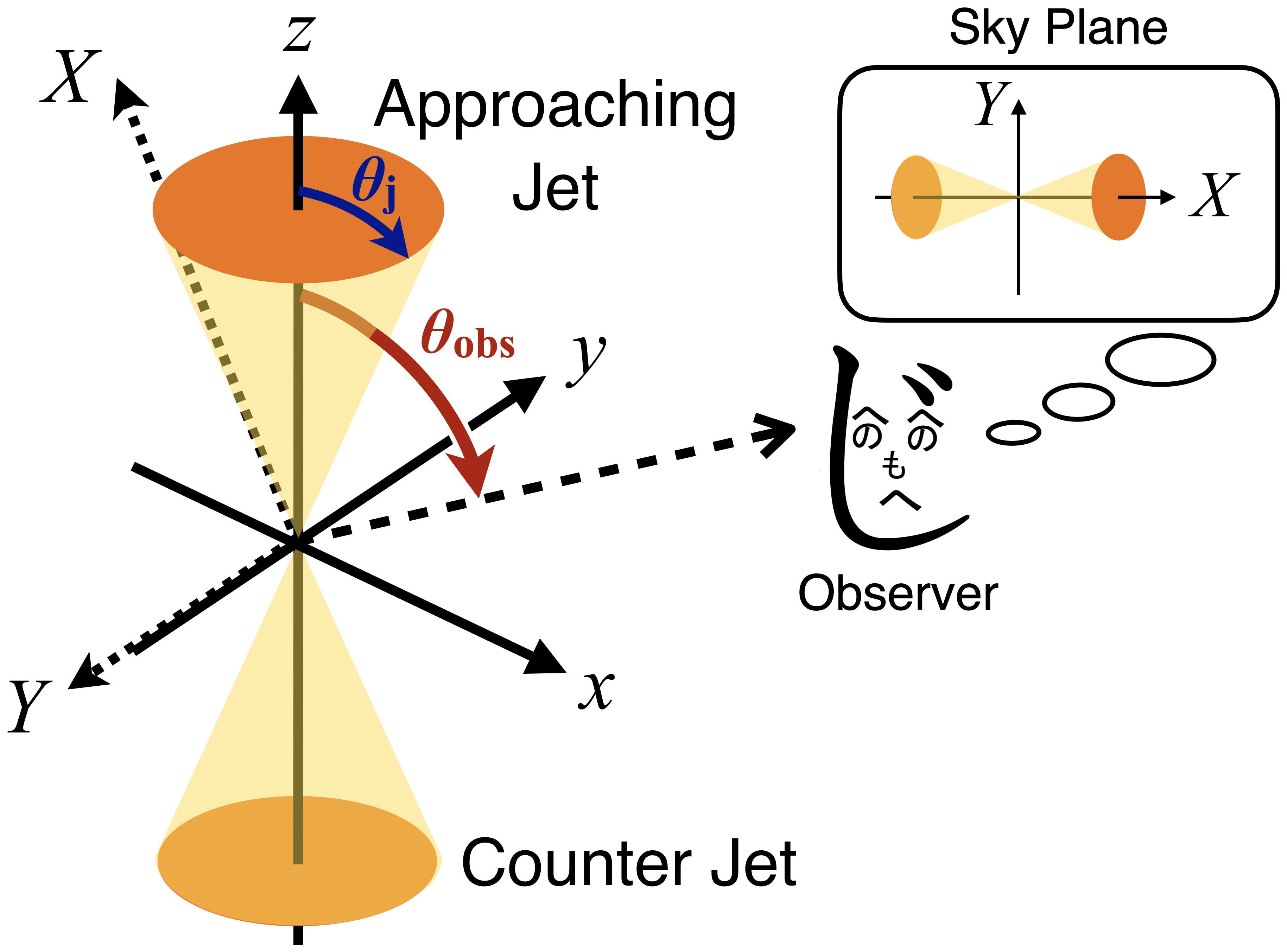}
    \caption{A schematic picture of the adopted coordinate system in our calculation. For a jet geometry, the jet is assumed to have a bipolar structure with a half-opening angle of $\theta_{\rm j}$, and each jet propagates along $z$ axis. An observer is located within the $xz$ plane, whose viewing angle $\theta_{\rm obs}$ is measured from the $z$ axis. The incoming (traveling into the positive $z$ direction) and outgoing jets are called an approaching and counter jets, respectively. The coordinate system of the sky ($XY$) plane is defined so that the approaching jet always propagates into the positive $X$ direction.}
    \label{fig:picture}
\end{figure}

\subsection{Dynamics}
The dynamics of the outflow is calculated under the piecewise spherical approximation. This is justified when the outflow is highly relativistic. However, for a jetted outflow, when its Lorentz factor becomes smaller than the inverse of the half opening angle, the jet starts to expand laterally \citep[e.g.,][]{Rhoads1999,Granot&Piran2012}. We neglect this sideway expansion in this paper, and assume that the jet propagates keeping its original shape. For TDE jets, this might be allowed because some jetted TDEs do not show signatures of the lateral expansion \citep{Berger+2012,Matsumoto&Metzger2023}.

We calculate the outflow's dynamics based on the energy conservation \citep{Piran+2013}, which is adopted in our previous studies \citep{Ricci+2021,Bruni+2021,Ho+2025}. For a top-hat jet, which has uniform energy and Lorentz factor distribution over its solid angle, the dynamics is characterized by the initial kinetic energy $E_{\rm ej}$ and Lorentz factor $\Gamma_0$, and the CNM density profile $n(r)$. Note here $E_{\rm ej}$ is the isotropic-equivalent energy, defined as the energy the outflow would have if it covered a solid angle of $4\pi$. Given the initial energy and Lorentz factor, the (isotropic-equivalent) mass of the outflow is given by the relation:
\begin{align}
E_{\rm ej}=(\Gamma_0-1)M_{\rm ej}c^2\ ,
    \label{eq:Eej initial}
\end{align}
where $c$ is the speed of light. The outflow expands sweeping up the CNM and unless the radiative energy loss is significant, its total energy at a radius $R$ is given by
\begin{align}
E_{\rm ej}=(\Gamma-1)[M_{\rm ej}+\Gamma M(R)]c^2\ ,
    \label{eq:Eej}
\end{align}
where the swept-up CNM mass is 
\begin{align}
M(R)&=\int_{}^{R}4\pi r^2 m_{\rm p} n(r) dr\ .
    \label{eq:Mswept}
\end{align}
Here $m_{\rm p}$ is the proton mass. Eq.~\eqref{eq:Eej} gives the outflow velocity, which is used to obtain the radius:
\begin{align}
R(t)=\int_{}^{t} \beta(t)cdt\ ,
\end{align}
where $t$ is the lab-frame time.

\subsection{Flux}
\label{sec:flux}
Once the dynamics is determined, the observed flux is calculated by integrating the emmisivity over the emitting volume. The following method is largely based on \cite{vanEerten+2010,Takahashi+2022} but we account for the cosmological effect explicitly.

The flux is formally given by the integration of the intensity over the solid angle subtended by an emitting source \citep{Rybicki&Lightman1979}:
\begin{align}
F_{\nu_{\rm obs}}=\int I_{\nu_{\rm obs}}d\widehat{\Omega}\ ,
	\label{eq:Fnu}
\end{align}
where $I_{\nu}$ is the intensity and $\nu_{\rm obs}$ is the observed frequency. The intensity is given by the radiative transfer equation:
\begin{align}
\frac{dI_{\nu}}{ds}=-\alpha_{\nu}I_{\nu}+j_{\nu}\ ,
    \label{eq:Inu}
\end{align}
where $\nu$ is the lab-frame frequency and related to $\nu_{\rm obs}$ by $\nu_{\rm obs}=\nu/(1+z)$. Under the thin-shell approximation, the absorption coefficient $\alpha_\nu$ and emissivity $j_\nu$ may be assumed to be constant, and one can obtain
\begin{align}
I_{\nu}=\frac{j_{\nu}}{\alpha_{\nu}}(1-e^{-\tau_{\nu}}){\,\,\,\rm and\,\,\,}\tau_{\nu}=\int \alpha_{\nu}ds=\alpha_{\nu}\Delta s\ ,
\end{align}
where $\tau_{\nu}$ and $\Delta s$ are the optical depth and the shell width along the direction to the observer, respectively. From the Lorentz invariance of $I_\nu/\nu^3$, the observed intensity is given by $I_{\nu_{\rm obs}}=I_{\nu}/(1+z)^3$. Plunging above equations into Eq.~\eqref{eq:Fnu}, one obtains 
\begin{align}
F_{\nu_{\rm obs}}&=\frac{(1+z)}{d_{\rm L}^{2}}\int j_{\nu}\left(\frac{1-e^{-\tau_{\nu}}}{\tau_{\nu}}\right)dV\ .
	\label{eq:Fnu2}
\end{align}
where $dV=dS\Delta s$ is the volume element and $dS$ is the emitting surface area projected on the sky plane. The latter is related to the solid angle by $d\widehat{\Omega}=dS/d_{\rm A}^2$. Here $d_{\rm L}$ and $d_{\rm A}=d_{\rm L}/(1+z)^2$ are the luminosity and angular diameter distances, respectively.

The absorption coefficient and emissivity are evaluated in the fluid rest frame. Quantities at the lab and fluid rest frames are related by the Lorentz transformation: $\nu=\delta_{\rm D}\nu^{\prime}$, $\alpha_{\nu}=\delta_{\rm D}^{-1}\alpha_{\nu^{\prime}}^{\prime}$, and $j_{\nu}=\delta_{\rm D}^{2}j_{\nu^{\prime}}^{\prime}$, where the quantities in the fluid rest frame are denoted with a prime, and the relativistic Doppler factor is defined by
\begin{align}
\delta_{\rm D}\equiv\frac{1}{\Gamma(1-\beta\mu)}\ .
    \label{eq:delta}
\end{align}
Here $\Gamma$ and $\beta$ are the Lorentz factor and normalized velocity of the fluid element, respectively, and $\mu$ is the cosine of the angle between the direction of the fluid motion and the observer’s line of sight in the lab frame. The expressions of $\alpha^\prime_{\nu^\prime}$ and $j^\prime_{\nu^\prime}$ are given in Appendix~\ref{sec:appendix}.

We introduce the spherical coordinate whose polar axis is aligned to the jet axis (see Fig.~\ref{fig:picture}), and rewrite the volume element by $dV=R^{2}\Delta Rd\Omega$. Here $\Delta R$ is the shell width along the radial direction, which is given by the conservation of the particle number \citep{vanEerten+2010,Takahashi+2022}:
\begin{align}
\Delta R=\frac{R}{4(3-k)\Gamma^{2}(1-\beta\mu)}\ .
	\label{eq:shell dR}
\end{align}
Here we consider the case where the CNM has a power-law radial dependence, $n(r)\propto r^{-k}$. The shell widths along the different directions are related by $\Delta s=\Delta R/\mu$.

The integration of Eq.~\eqref{eq:Fnu2} should be carried out by taking into account the light traveling effect. A photon emitted at lab time $t$ from a fluid element with $\mu$ is received by the observer at observer time of $T$ when the following condition is satisfied:
\begin{align}
\frac{T}{1+z}=t-\frac{\mu R(t)}{c}\ .
	\label{eq:EATS}
\end{align}
For given observer time $T$, the set of radii $R(t)$ obtained for different $\mu$ forms a surface, commonly referred to as the equal-arrival-time surface (EATS). Note that in the coordinate system in Fig.~\ref{fig:picture}, $\mu$ is simply given by 
\begin{align}
\mu=\vb*{n}_{\rm obs}\cdot\vb*{e}_{r}=\cos\varphi\sin\theta\sin\theta_{\rm obs}+\cos\theta\cos\theta_{\rm obs}\ ,
    \nonumber
\end{align}
where 
\begin{align}
\vb*{n}_{\rm obs}&=(\sin\theta_{\rm obs},0,\cos\theta_{\rm obs})\ ,
	\nonumber\\
\vb*{e}_{r}&=(\cos\varphi\sin\theta,\sin\varphi\sin\theta,\cos\theta)\ .
    \nonumber
\end{align}

\subsection{Image}
To describe the image, we introduce a new ($X,Y$) coordinate as shown in Fig.~\ref{fig:picture}. The $X$ axis is defined so that it coincides with the axis of an approaching jet. The basis vectors of this coordinate are given by
\begin{align}
\vb*{e}_{X}&=(-\cos\theta_{\rm obs},0,\sin\theta_{\rm obs})\ ,
    \nonumber\\
\vb*{e}_{Y}&=(0,-1,0)\ ,
    \nonumber
\end{align}
and a fluid element at $\vb*{R}=R\vb*{e}_{r}$ is projected to
\begin{align}
X&=\vb*{R}\cdot\vb*{e}_{X}=R(-\cos\varphi\sin\theta\cos\theta_{\rm obs}+\cos\theta\sin\theta_{\rm obs})\ ,
	\nonumber\\
Y&=\vb*{R}\cdot\vb*{e}_{Y}=-R\sin\varphi\sin\theta\ .
    \nonumber
\end{align}
On the sky plane, a corresponding angular size is given by $\widehat{X}=X/d_{\rm A}$.

The flux is formally given by 
\begin{align}
F_{\nu_{\rm obs}}=\int\frac{d^{2}F_{\nu_{\rm obs}}}{dXdY}dXdY=\int\frac{d^{2}F_{\nu_{\rm obs}}}{dXdY}\frac{\partial(X,Y)}{\partial(\theta,\varphi)}d\theta d\varphi\ .
	\label{eq:dF/dXdY}
\end{align}
Here the Jacobian is given by 
\begin{align}
&\frac{\partial(X,Y)}{\partial(\theta,\varphi)}=|\vb*{N}\cdot\vb*{n}_{\rm obs}|
	\label{eq:jacobian}\\
&\,\,\,\,\,\,\,\,\,=R\frac{\partial R}{\partial \theta}(\cos\theta_{\rm obs}\sin^2\theta-\sin\theta_{\rm obs}\sin\theta\cos\theta\cos\varphi)
    \nonumber\\
&\,\,\,\,\,\,\,\,\,\,\,\,\,\,\,+R\frac{\partial R}{\partial \varphi}\sin\theta_{\rm obs}\sin\varphi
    \nonumber\\
&\,\,\,\,\,\,\,\,\,\,\,\,\,\,\,+R^2(\cos\theta_{\rm obs}\sin\theta\cos\theta+\sin\theta_{\rm obs}\sin^2\theta\cos\varphi)\ ,
    \nonumber
\end{align}
and $\vb*{N}$ is the normal vector to the EATS given by
\begin{align}
\vb*{N}=-R\sin\theta\frac{\partial R}{\partial \theta}\frac{\partial\vb*{e}_{r}}{\partial \theta}-\frac{R}{\sin\theta}\frac{\partial R}{\partial \varphi}\frac{\partial \vb*{e}_r}{\partial \varphi}+R^2\sin\theta\vb*{e}_{r}\ .
\end{align}
Note $R$ denotes the radius of the EATS, $R_{\rm EATS}(\theta,\varphi)$
given by Eq.~\eqref{eq:EATS}. By comparing Eqs.~\eqref{eq:Fnu2} and \eqref{eq:dF/dXdY}, one can obtain the expression of the intensity:
\begin{align}
I_{\nu_{\rm obs}}=\frac{d^{2}F_{\nu_{\rm obs}}}{d\widehat{X}d\widehat{Y}}=\frac{\delta_{\rm D}^{2}j_{\nu^{\prime}}^{\prime}}{(1+z)^{3}}\left(\frac{1-e^{-\tau_{\nu}}}{\tau_{\nu}}\right)\frac{R^{2}\Delta R\sin\theta}{|\vb*{N}\cdot\vb*{n}_{\rm obs}|}\ .
	\label{eq:dF/dXdY2}
\end{align}
Once an intensity distribution is obtained, the location of the emission centroid on the sky is calculated by
\begin{align}
(\widehat{X}_{\rm cen},\,\widehat{Y}_{\rm cen}) =\frac{\int (\widehat{X},\,\widehat{Y})I_{\nu_{\rm obs}}d\widehat{X}d\widehat{Y}}{\int I_{\nu_{\rm obs}}d\widehat{X}d\widehat{Y}}\ .
	\label{eq:Xcen}
\end{align}
In particular, for an axisymmetric jet considered in this study, the emission centroid is always on the $X$-axis ($\widehat{Y}_{\rm cen}=0$).

\section{Image of TDE Radio Flare: Application to AT2018hyz}\label{sec:result}

We calculate images of TDE radio flares by using the method in the previous section. Our main focus is on investigating the origin of late-time radio flares in TDEs, in particular, AT2018hyz showing one of the brightest radio flares \citep{Cendes+2022b,Sfaradi+2024,Cendes+2026}. AT2018hyz was discovered as an optical TDE, having a typical time evolution with a peak bolometric luminosity of $\simeq 10^{44}\,\mathrm{erg\,s^{-1}}$ \citep{Gomez+2020, Short+2020, vanVelzen+2021, Hammerstein+2023}. A prompt radio followup at $\simeq30\,\rm days$ reported no detection, but a long-term monitoring revealed a sudden brightening $\sim1000\,\rm days$ after discovery 
\citep{Cendes+2022b, Sfaradi+2024, Cendes+2026}. 
The radio light curve shows a rapid rise $F_{\nu} \propto T^{5}$ and the spectra are fitted by a single-peak spectrum implying a synchrotron self-absorption (SSA) signature. The recent observation at $\simeq2000\,\rm days$ confirms that the light curve is still rising but at a little slower pase $\propto T^{3}$, and its luminosity is comparable to jetted TDEs \citep{Cendes+2026}.

Such an onset of the radio flare as late as $\sim1000\,\rm days$ was totally unexpected, and several scenarios have been proposed. One possibility is that the radio emitting outflow was launched after a delay of $\sim1000\,\rm days$ since the optical flare. This scenario is motivated by the result of the equipartition analysis \citep{BarniolDuran+2013,Cendes+2022b}. The inferred radial evolution suggests that the outflow was launched not at the time of optical discovery, but rather $\simeq 700\,\rm days$ later with a Newtonian velocity \citep{Cendes+2022b,Cendes+2026}. In this case, a state transition of an accretion disk may be responsible for launching the outflow with the significant delay \citep{Alexander+2026}. In other scenario, the radio outflow is a relativistic jet launched around the optical discovery but toward a different direction from our line of sight. Due to the off-axis effect, the radio flux is initially suppressed until it decelerates. This possibility is pointed out by \citet{Matsumoto&Piran2023}, who generalized the equipartition method to arbitrary viewing angles. Subsequent forward-modeling studies also confirmed this scenario \citep{Sfaradi+2024,Sato+2024,Cendes+2026}. There are also other scenarios such as an interaction between a prompt-launched outflow and a gas cloud at some distance from the galactic center \citep[e.g.,][]{Zhuang+2025}, but we focus on the scenarios of outflows interacting with CNM.

The current observations with radio light curve and spectra cannot discriminate the delayed outflow and off-axis jet scenarios, and they are still remain viable explanations \citep{Cendes+2026}. The most direct way to distinguish between them is through radio imaging with VLBI. Motivated by this, we compute radio images for both scenarios and quantitatively examine the differences between them.

\subsection{Light Curve \& Spectrum}
\label{sec:light curve and spectra}

We begin by fitting the light curve and spectrum of AT2018hyz, which are calculated by using the method outlined in Section~\ref{sec:flux}, to constrain the model parameters for both scenarios. While the parameters can be estimated using the equipartition analysis, this method does not account for the temporal evolution. As a result, the inferred parameters do not necessarily reproduce the observed time-dependent behavior. We therefore independently determine parameter sets based on our forward modeling, although they do not significantly deviate from those inferred from the equipartition analysis. In this paper, we do not attempt to determine the best-fitting parameters; instead, we search for parameter sets that reasonably reproduce the observations.

\subsubsection{Delayed Outflow Scenario}
\label{sec:delayed outflow}

\begin{figure}
    \centering
    \includegraphics[width=85mm,bb=0 0 285 219]{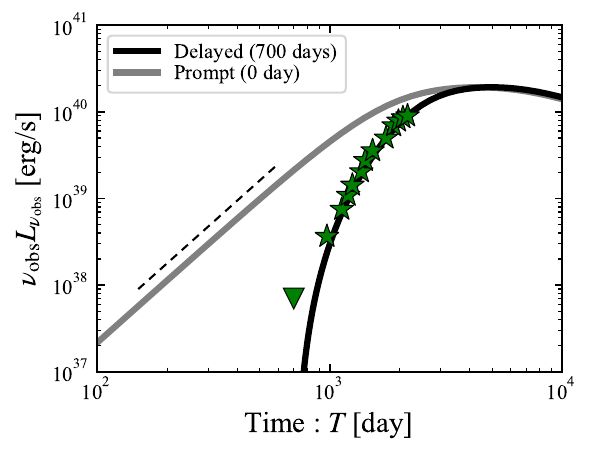}
    \includegraphics[width=85mm,bb=0 0 272 185]{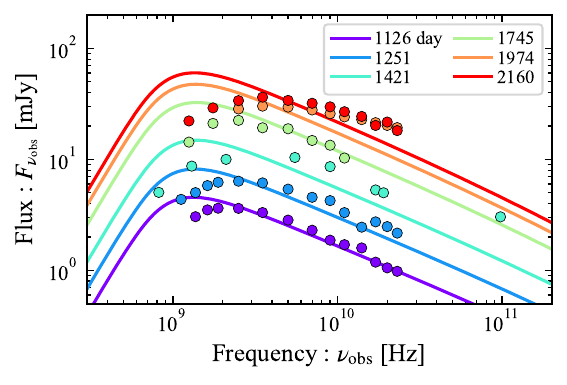}
    \caption{The radio light curve and spectrum of AT2018hyz for the delayed outflow model. ({\bf Top}) The $5\,\rm GHz$ light curve. The black and gray curves show the model light curves of a spherical outflow launched at $T=700$ (delayed) and $0\,\rm day$ (prompt), respectively. The black dashed line represents the analytical scaling of the flux, $F_\nu\propto T^{\frac{12-k(p+5)}{4}}$. ({\bf Bottom}) The radio spectrum at each epoch. Solid curves show the model spectra corresponding to the black curve in the top panel.}
    \label{fig:18hyz_sph}
\end{figure}

Figure~\ref{fig:18hyz_sph} depicts the radio light curve at 5 GHz and spectra for different epochs. The adopted parameters are shown in Table~\ref{table:18hyz_para}, and they are obtained by manually varying each value around those inferred from the equipartition analysis to give reasonable fits. We assume that the outflow is spherical and launched at $700\,\rm days$ since the optical discovery as suggested by the previous studies \citep{Cendes+2022b,Cendes+2026}. Although our calculations are based on numerical integration over the jet surface, the results can be largely described by analytical expressions \citep{Matsumoto&Piran2021b,Matsumoto&Piran2024}.\footnote{In the following analytical argument, we ignore the negligible redshift effect for AT2018hyz ($z=0.0457$). In particular, for the Newtonian delayed outflow the lab and observer time are effectively identical, $t\simeq T$.}

The sharp rise in the light curve is caused by the offset in the origin of time. In this phase, the 5 GHz band is optically thin (see the bottom panel of Fig.~\ref{fig:18hyz_sph}), and hence the increase in flux results from accumulation of emitting electrons swept up by the freely coasting outflow, which is given by $F_{\nu}\propto R^{\frac{12-k(p+5)}{4}}$ (Eq.~\ref{eq:Fnu2}). The gray curve shows the light curve from a promptly launched outflow, which agrees with the above scaling. Therefore for given $p$ and $T_0$, the light-curve slope constrains the slope of the density profile, $k$. The flux peaks when the outflow begins decelerating.

In contrast to the light curve, obtaining reasonable spectral fits is challenging, particularly in the optically thick, low-frequency regime. Observations find the peak frequency staying around $\simeq3\,\rm GHz$ \citep{Cendes+2026}, whereas our result gives a slightly lower value of $\simeq 1\,\rm GHz$. While the peak is rounded, it could be interpreted as being shaped by SSA whose peak flux $F_{\nu_{\rm a}}$ and frequency $\nu_{\rm a}$ are given by \citep{Matsumoto&Piran2021b}
\begin{align}
F_{\nu_{\rm a}}&\propto\left[\varepsilon_{\rm e}^{5(p-1)}\varepsilon_{\rm B}^{\frac{2p+3}{2}}n^{\frac{2p+13}{2}}R^{2p+13}\beta^{\frac{12p-7}{2}}\right]^{\frac{1}{p+4}}
    \label{eq:F peak}\\
&\overset{p=2.2}{\propto}\varepsilon_{\rm e}^{0.97}\varepsilon_{\rm B}^{0.60}n^{1.4}\beta^{4.4}\ ,
    \nonumber\\
\nu_{\rm a}&\propto\left[\varepsilon_{\rm e}^{2(p-1)}\varepsilon_{\rm B}^{\frac{p+2}{2}}n^{\frac{p+6}{2}}R^{2}\beta^{5p-2}\right]^{\frac{1}{p+4}}
    \label{eq:nu peak}\\
&\overset{p=2.2}{\propto}\varepsilon_{\rm e}^{0.39}\varepsilon_{\rm B}^{0.34}n^{0.66}\beta^{1.8}\ .
    \nonumber
\end{align}
Both flux and frequency increase with the parameters, but the flux exhibits a stronger dependence than the frequency. This makes it difficult to increase the frequency by naively changing the parameters with fixing the flux. Since detailed fitting is not the purpose of this paper and the SSA frequency also depends on the outflow geometry, we do not pursue the discrepancy further.

Our resulting energy $E_{\rm ej}=10^{52}\,\rm erg$ is $\simeq50$ times larger than the equipartition estimate by \cite{Cendes+2026}. However, the discrepancy is not a serious issue because the equipartition analysis measures only the energy in the shocked and radio emitting region, which can be smaller for freely coasting outflows \citep{Matsumoto+2022}. In addition, such a large energy can also be inferred from the observational indication that the outflow remains in a free-expansion phase at the latest observation, $\simeq2000\,\rm days$. Assuming a constant CNM profile, the deceleration time is given by
\begin{align}
T_{\rm dec}\simeq\left(\frac{3E_{\rm ej}}{2\pi m_{\rm p}c^{5} n \beta_{0}^{5}}\right)^{1/3}\ .
    \label{eq:t_dec}
\end{align}
Since the deceleration effect appears a little earlier than $T_{\rm dec}$, say half of the timescale, we may infer the timescale longer than $\simeq2\times(2000-700)=2600\,\rm days$, where we account for the shift of $700\,\rm days$ in the time origin. Then we obtain a lower limit on the energy by $T_{\rm dec}\gtrsim2600+700=3300\,\rm days$, as
\begin{align}
E_{\rm ej}\gtrsim5.8\times10^{51}\,{\rm erg}\left(\frac{n}{1\,{\rm cm^{-3}}}\right)\left(\frac{\beta_{0}}{0.3}\right)^{5}\left(\frac{T}{3300\,{\rm day}}\right)^{3}\ .
\end{align}
Here the density is motivated by our resulting profile and the size of the outflow.

We comment on the value of our $\varepsilon_{\rm B}=0.1$ larger than that of \cite{Cendes+2026}, $\sim10^{-3}$, who used an X-ray upper limit. Our synchrotron flux does not violate the limit. This is because we fix $p=2.2$ for entire our calculation while \cite{Cendes+2026} adopt $p<2$, which is inferred by their spectral fit.

\begin{table*}
\centering
\caption{Model Parameters of AT2018hyz for the Delayed Outflow and Off-axis Jet Scenarios.}
\label{table:18hyz_para}
\begin{tabular}{clcc}
\hline
Symbol&Parameter&Delayed Outflow&Off-axis Jet\\
\hline
$\theta_{\rm j}$&Half opening angle&$\pi$&0.1 rad\\
$\beta_{0}$&Initial velocity&$0.25$&$0.999^{\rm (a)}$\\
$\Gamma_{0}$&Initial Lorentz factor&$1.033^{\rm (a)}$&$30$\\
$E_{\rm ej}$&Kinetic energy&$10^{52}$ erg&$10^{56}$ erg\\
$\widetilde{n}$&Normalization of CNM density profile at $3\times10^{17}\,\rm cm$&$8\,\rm cm^{-3}$&$2\,\rm cm^{-3}$\\
$k$&Slope of CNM density profile&$1/3$&$0$\\
$p$&Slope of electron distribution&$2.2$&$2.2$\\
$\varepsilon_{\rm e}$&Energy fraction of non-thermal electron energy&$0.3$&0.07\\
$\varepsilon_{\rm B}$&Energy fraction of magnetic field &$0.1$&0.001\\
\hline
$T_{0}$&Outflow launch time&$700$ day&0 day\\
\hline
$\theta_{\rm obs}$&Best-fit viewing angle&-&$70^\circ$\\
\hline
\multicolumn{4}{l}{$^{\rm (a)}$ Derived from the corresponding velocity (Lorentz factor) and included for completeness.}
\end{tabular}
\end{table*}

\subsubsection{Off-axis Jet Scenario}
\label{sec:jet}

\begin{figure}
    \centering
    \includegraphics[width=85mm,bb=0 0 277 214]{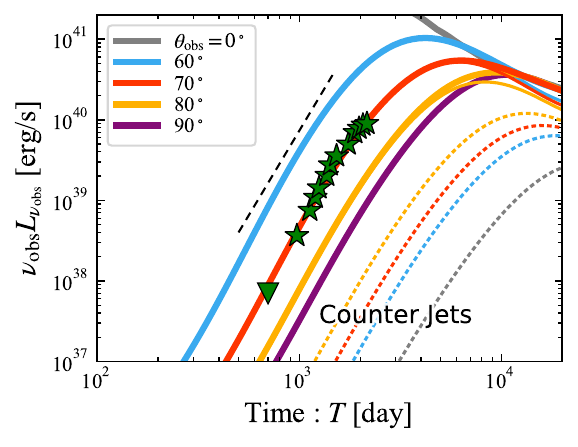}
    \includegraphics[width=85mm,bb=0 0 272 185]{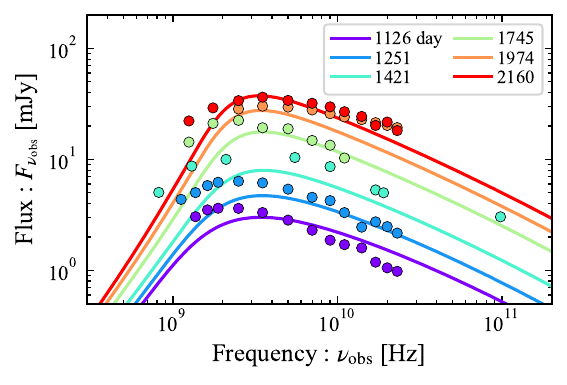}
    \caption{The same as Fig.~\ref{fig:18hyz_sph} but for the off-axis jet model with different viewing angles. ({\bf Top}) The thick solid curves show the total radio luminosity contributed by both approaching (thin solid) and counter (dotted) jets while the former almost overlaps with the thick solid curves for $\lesssim10000\,\rm days$. The black dashed line represents the analytical scaling, $F_\nu\propto T^{\frac{3(5-p)}{2}}$. ({\bf Bottom}) The solid curves denote the model spectra corresponding to $\theta_{\rm obs}=70^\circ$.}
    \label{fig:18hyz_jet}
\end{figure}

Figure~\ref{fig:18hyz_jet} depicts the radio light curve and spectrum calculated for the off-axis jet model. The adopted parameters are chosen in the same way as in the delayed outflow model and shown in Table~\ref{table:18hyz_para}. The light curves rise rapidly as the jet decelerates and weakens the off-axis effect. The radio flux is larger for smaller viewing angles, and the contribution of the counter jet is completely negligible unless $\theta_{\rm obs}\gtrsim80^\circ$, until the light curves peak. The peak is reached when the jet becomes Newtonian $\Gamma\simeq\delta_{\rm D}\simeq1$.

The evolution in the rising phase can be understood analytically. For a viewing angle larger than $\theta_{\rm obs}\gg\theta_{\rm j}$ and $1/\Gamma$, the flux is contributed almost equally by the entire jet, so that the one-zone approximation is applicable. The optically-thin synchrotron flux is given from Eq.~\eqref{eq:Fnu2}:
\begin{align}
F_{\nu_{\rm obs}}&\propto\varepsilon_{\rm e}^{p-1}\varepsilon_{\rm B}^{\frac{p+1}{4}} n^{\frac{p+5}{4}}R^{3}\theta_{\rm j}^{2}\Gamma^{\frac{3p-1}{2}}\delta_{\rm D}^{\frac{p+5}{2}}\ .
    \label{eq:Fnu rise}
\end{align}
For a relativistic outflow with $R(t)\simeq ct$ viewed from off axis $\mu\ll1$, the radius is given by (Eq.~\ref{eq:EATS})
\begin{align}
R\simeq\frac{cT}{1-\mu}\ .
    \label{eq:R}
\end{align}
Note that the outflow apparently does not decelerate ($R\propto T$) in the off-axis regime even in the deceleration phase with $\Gamma\propto(E_{\rm ej}/nR^3)^{1/2}$. We take a point on the jet axis as a representative point (hereafter called the jet center), and set $\mu=\cos\theta_{\rm obs}$. By approximating\footnote{We use $1-\cos\theta_{\rm obs}\simeq\theta_{\rm obs}^2/2$, which remains accurate to within 20\% even at $\theta = \pi/2$.} the Doppler factor (Eq.~\ref{eq:delta}) and radius (Eq.~\ref{eq:R}) as $\delta_{\rm D}\simeq2/\Gamma\theta_{\rm obs}^2$ and $R\simeq2cT/\theta_{\rm obs}^2$, and plunging these expressions into Eq.~\eqref{eq:Fnu rise} we obtain
\begin{align}
F_{\nu_{\rm obs}}&\propto\varepsilon_{\rm e}^{p-1}\varepsilon_{\rm B}^{\frac{p+1}{4}}n^{\frac{11-p}{4}}T^{\frac{3(5-p)}{2}}E_{\rm ej}^{\frac{p-3}{2}}\theta_{\rm j}^{2}\theta_{\rm obs}^{2(p-10)}
    \label{eq:Fnu jet2}\\
&\overset{p=2.2}{\propto}\varepsilon_{\rm e}^{1.2}\varepsilon_{\rm B}^{0.8}n^{2.2}E_{\rm ej}^{-0.4}T^{4.2}\theta_{\rm j}^{2}\theta_{\rm obs}^{-15.6}\ .
    \nonumber
\end{align}
The scaling of $T$ agrees with our numerical results \citep[see also][]{Beniamini+2023b}. In this example, $\theta_{\rm obs} = 70^{\circ}$ provides the best match to the observations, although there is a degeneracy among parameters as seen in Eq.~\eqref{eq:Fnu jet2}. 

The bottom panel of Fig.~\ref{fig:18hyz_jet} shows the spectra for $\theta_{\rm obs}=70^{\circ}$. The quality of the fit seems better than the delayed outflow model (Fig.~\ref{fig:18hyz_sph}), and in particular the spectral peaks are located at $\simeq 3\,\rm GHz$. The early-time model spectra ($\lesssim1400\,\rm days$) exhibit a rounded peak in contrast to the later ones. This is because the SSA and synchrotron characteristic frequencies (see Appendix~\ref{sec:appendix} for the latter's definition), overlap at the early phase, $\nu_{\rm a}\sim\nu_{\rm m}$, which is also implied by the equipartition analysis \citep{Cendes+2026}. $\nu_{\rm m}$ decreases faster, resulting in a narrow peak shaped only by SSA.

Our model suggests a huge jet energy, $E_{\rm j,iso}=E_{\rm ej}\sim10^{56}\,\rm erg$, even after accounting for the beaming correction, $E_{\rm j} \sim \theta_{\rm j}^{2} E_{\rm j,iso} \sim 10^{54}\,\rm erg$.
This is about two orders of magnitude larger than the energies typically inferred for other jetted TDEs \citep[e.g.,][]{Berger+2012,Matsumoto&Metzger2023}. Although this energy is exceptionally large, it is required by the observation that the jet remains relativistic at $\simeq2000\,\rm days$, while jets in other jetted TDEs have already decelerated to the Newtonian regime by similar epochs. Practically this large energy comes from two conditions. One is that the jet should be decelerating at the detection of the radio flare, $T\sim1000\,\rm day$:
\begin{align}
E_{\rm ej}&\lesssim\frac{4\pi}{3}m_{\rm p}c^{2}n\Gamma^{2}R^{3}\simeq8.8\times 10^{55}\,{\rm erg}
    \nonumber\\
&\times\left(\frac{n}{1{\,\rm cm^{-3}}}\right)\left(\frac{\Gamma}{10}\right)^2\left(\frac{T}{1000\,\rm day}\right)^{3}\theta_{\rm obs}^{-6}\ ,
    \label{eq:condition Eej jet}
\end{align}
where we used $R\simeq2cT/\theta_{\rm obs}^2$. The other comes from the fact that the jet should be still relativistic at the latest observation $T\simeq2000\,\rm day$, since the observed light curve is still rising and has not yet reached its peak. This sets a lower limit on the jet energy:
\begin{align}
E_{\rm ej}&\gtrsim\frac{4 \pi}{3} m_{\rm p} c^{2} n R^{3}\simeq 7.0 \times 10^{54}\,{\rm erg}
    \nonumber\\
&\times\left(\frac{n}{1{\,\rm cm^{-3}}}\right) \left(\frac{T}{2000\,\rm day}\right)^{3}\theta_{\rm obs}^{-6}\ .
    \label{eq:condition Eej jet2}
\end{align}

The inferred energy can likely be reduced by at most about an order of magnitude through parameter degeneracies, but not to values comparable to those of other jetted TDEs. The main constraint comes from Eq.~\eqref{eq:condition Eej jet2}: reducing $E_{\rm ej}$ leads to an earlier transition to the Newtonian phase and hence an earlier light-curve peak. This can be compensated only by reducing $n$ by a similar factor, which in turn lowers the radio flux (Eq.~\ref{eq:Fnu jet2}). Restoring the flux requires increasing $\varepsilon_{\rm e}$ and $\varepsilon_{\rm B}$, and/or decreasing $\theta_{\rm obs}$. Such adjustments are, however, limited because the former shifts $\nu_{\rm m}$ to higher frequencies and worsens the fit to the early-time observed spectrum, and the latter causes the jet to enter the Newtonian phase at an earlier time. We find that, under these constraints, the energy can be reduced by at most a factor of $\sim10$. A similar argument applies to the delayed-outflow model, for which the inferred energy is likewise robust against parameter degeneracies.

\subsection{Image}
\label{sec:image}

\begin{figure}
    \centering
    \includegraphics[width=85mm,bb=0 0 303 400]{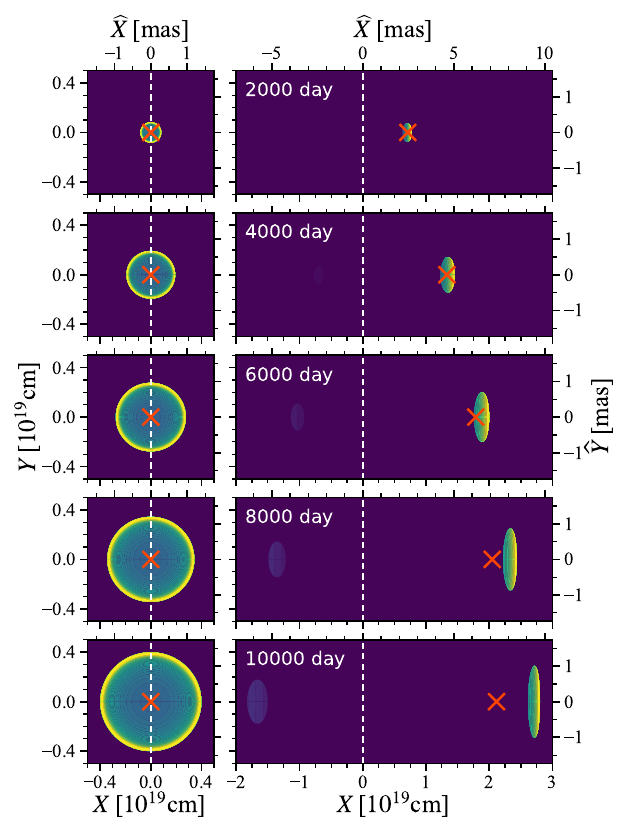}
    \caption{The synthetic radio intensity maps of AT2018hyz at 5 GHz for the delayed outflow (left) and off-axis jet models ($\theta_{\rm obs}=70^\circ$, right). In each panel, the left and bottom axes represent the physical distance, while the right and top ones represent the corresponding angular distance. The red crosses show the location of the combined emission centroid by approaching and counter jets.}
    \label{fig:18hyz_im}
\end{figure}

Figure~\ref{fig:18hyz_im} shows the synthetic radio intensity maps at 5 GHz calculated for the delayed outflow (left) and off-axis jet (right) scenarios. The parameters are the same as ones used in the calculations for the light curves and spectra. In each map, the surface brightness is normalized to the upper 95\% of the distribution, and values exceeding the 95th percentile are clipped to that percentile level. For AT2018hyz with $z=0.0457$, the angular diameter distance is $d_{\rm A}\simeq192.4\,\rm Mpc$,\footnote{A flat $\Lambda$CDM cosmology with $H_0 = 67.7\,\rm km \,s^{-1}\, Mpc^{-1}$, $\Omega_{\rm m} = 0.315$, and $\Omega_\Lambda = 0.685$ is adopted \citep{PlanckCollaboration2020_parameter}.} and hence the angular size is $\widehat{X}\simeq3.5{\,\rm mas\,}(X/10^{19}\,\rm cm)$. At 5 GHz, the entire emitting region in both scenarios is optically thin.

The left column of Fig.~\ref{fig:18hyz_im} shows the time evolution of the radio image for the delayed outflow. With the spherical geometry, the outflow appears disk-like on the sky and expands at a constant speed before the deceleration time. After $\simeq4000\,\rm days$, the deceleration becomes significant and the disk grows more slowly in size. Since the emissivity is distributed uniformly over the shock surface, the brightness distribution is described by
\begin{align}
I_{\nu_{\rm obs}}\propto\frac{1}{|\mu|} = \frac{1}{\sqrt{1-(\varpi/R)^2}}\ ,
\end{align}
where $\varpi^2=X^2+Y^2$. This explains the limb brightening of the images. For frequencies lower than $\nu_{\rm a}$, the brightness distribution is uniform since $\tau_{\nu}\propto 1/|\mu|$ (see Eq.~\ref{eq:dF/dXdY2}).

The right column of Fig.~\ref{fig:18hyz_im} shows the images of the off-axis jet. Initially the approaching jet dominates the emission and it moves toward $X>0$ on the sky. Around $\simeq6000\,\rm days$, the jet becomes Newtonian producing the light curve peak, and the counter jet emerges gradually after that. The intensity is higher for the right side of the approaching-jet image because a ray can intersect more fluid elements (same as the limb brightening of the delayed outflow). At optically thick frequencies, the brightness contrast across the image becomes weaker, qualitatively similar to the case of a Newtonian outflow. If the thin-shell approximation is relaxed and the radial structure of the outflow is taken into account, the image morphology would acquire a frequency dependence. In particular, the apparent image size is expected to become larger at lower frequencies \citep[e.g.,][]{Granot+1999b}.

\begin{figure}
    \centering
    \includegraphics[width=85mm,bb=0 0 313 214]{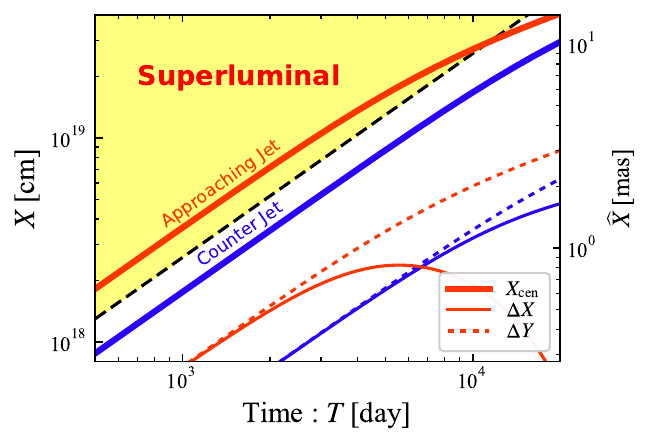}
    \caption{The evolution of the image position and size for the individual approaching (red) and counter (blue) jets for $\theta_{\rm obs}=70^\circ$. The thick solid curves shows the jet positions represented by the emission centroid. The counter-jet position is plotted with the sign reversed, $-X_{\rm cen}>0$. For the counter jet, The thin solid and dotted curves denote the width of each image in the $X$ and $Y$ directions, respectively. The yellow shaded region represents the distance accessible only for a superluminal emission source ($X_{\rm cen}\geq cT$).}
    \label{fig:18hyz_X}
\end{figure}

\begin{figure}
    \centering
    \includegraphics[width=85mm,bb=0 0 340 181]{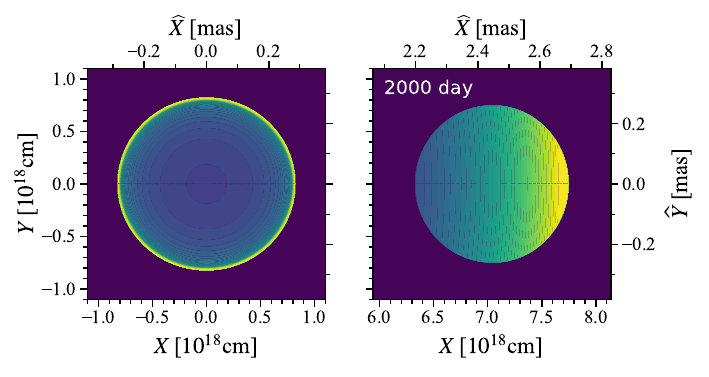}
    \caption{The same as Fig.~\ref{fig:18hyz_im} but the $X$ and $Y$-axes in both panels are set to equal scales, ensuring that circles are displayed without distortion. The right panel shows only the approaching jet. The intensity normalization in this figure is the upper 99\% of the distribution.}
    \label{fig:18hyz_im_early}
\end{figure}

Figure~\ref{fig:18hyz_X} shows the evolution of the image position and size for the individual jets. The image position is represented by the emission centroid given by Eq.~\eqref{eq:Xcen}. The centroids of both jets move at a constant velocity until deceleration sets in. Most importantly, the approaching jet exhibits superluminal motion with an apparent velocity of \citep[e.g.,][]{Rees1966}
\begin{align}
\beta_{\rm app}=\frac{\beta\sin\theta_{\rm obs}}{1-\beta\cos\theta_{\rm obs}}\ .
    \label{eq:beta_app}
\end{align}
Detection of such motion will be a smoking gun signature of a relativistic jet in delayed radio flares.

The jet images have an elliptical shape, whose width in the $X$ and $Y$ directions are also shown in Fig.~\ref{fig:18hyz_X}. Initially the images are nearly circular, but become gradually elongated in the $Y$ direction as seen in Fig.~\ref{fig:18hyz_im}. In particular, interestingly the jets in the relativistic phase have a filled-disk shape like the delayed outflow. In Fig.~\ref{fig:18hyz_im_early}, we show the images of both delayed outflow and approaching off-axis jet with equally scaled axes. While the shapes of their images are identical, the intensity distributions are different.

\begin{figure}
    \centering
    \includegraphics[width=85mm,bb=0 0 1443 1044]{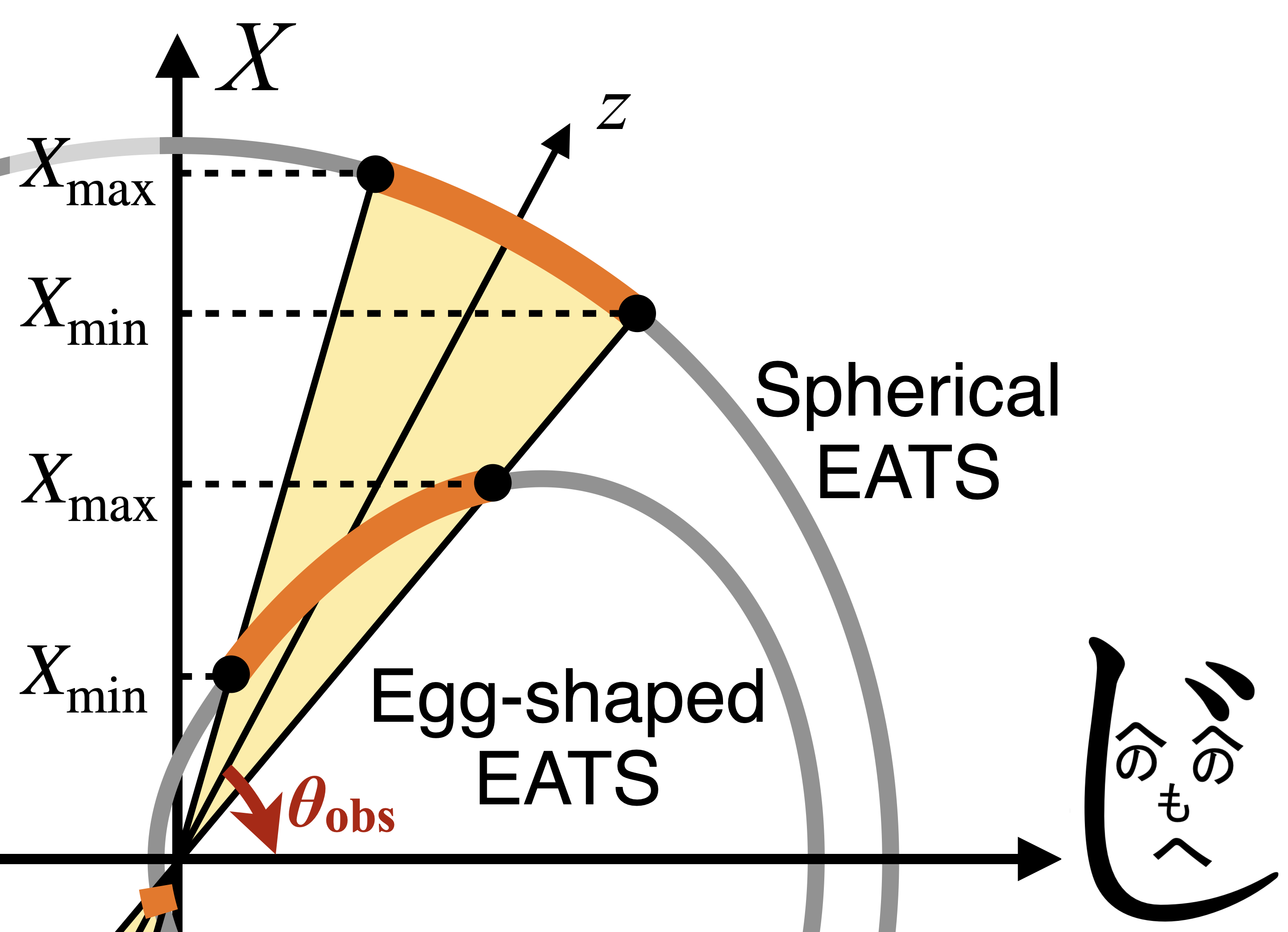}
    \caption{A schematic picture of the EATS of the off-axis jet at relativistic and Newtonian regimes. The EATS is given by the intersection of that of a spherical outflow (gray curves) and the jet (shaded region). In the relativistic phase, the former has an egg-like shape, and the far and near sides of the jet edge (black dots) are projected to smaller and larger $X$ coordinates on the sky plane, which we denote as $X_{\rm min}$ and $X_{\rm max}$, respectively. In the Newtonian phase, the far (near) side is projected to $X_{\rm max}$ ($X_{\rm min}$).}
    \label{fig:image_eats}
\end{figure}

The evolution of the individual jet images can be understood geometrically. The image position is approximately given by the projection of the jet center onto the sky plane, $X_{\rm j}\simeq R_{\rm j}\sin\theta_{\rm obs}$, where $R_{\rm j}$ is the radius of the jet center and given by Eq.~\eqref{eq:R} with $\mu=\cos\theta_{\rm obs}$. The width in the $Y$ direction is given by
\begin{align}
\Delta Y \simeq 2X_{\rm j}\sin\theta_{\rm j}\simeq2R_{\rm j}\theta_{\rm j}\ .
    \label{eq:dY}
\end{align}
To derive $\Delta X$, we have to consider the EATS of the jets, which is described by an intersection between an egg-shaped EATS of a spherically symmetric relativistic outflow \citep{Sari1998,Granot+1999}, and the jet geometry. Figure~\ref{fig:image_eats} illustrates the schematic picture of the EATS. In the relativistic phase, the far and near sides of the jet edge (black dots) are projected to smaller and larger X-coordinates, which we denote as $X_{\rm min}$ and $X_{\rm max}$, respectively. Their radii are also given by Eq.~\eqref{eq:R} with $\mu=\cos(\theta_{\rm obs}\pm\theta_{\rm j})$, where the plus (minus) sign corresponds to the far (near) side of the jet. Multiplying them by $\sin(\theta_{\rm obs}\pm\theta_{\rm j})$ to project on the sky plane, the width in the $X$ direction is given by
\begin{align}
\Delta X &= X_{\rm max} - X_{\rm min}\simeq 2R_{\rm j}\theta_{\rm j}\ ,
    \label{eq:dX}
\end{align}
which is comparable to $\Delta Y$ (Eq.~\ref{eq:dY}). As the jet gradually decelerates, Eq.~\eqref{eq:R} cannot capture the evolution of $R$, and the above scalings no longer apply (we will discuss the late-time evolution in the next section).

The angular resolution of global VLBI is typically $\sim\lambda_{\rm 5GHz}/10000{\,\rm km}\simeq\rm mas$, where $\lambda_{\rm 5GHz}=6\,\rm cm$. This implies that, for AT2018hyz, it may be difficult to resolve any substructure within each individual jet image unless the source is bright enough. Nevertheless, the approaching and counter jets are expected to be separated after a few thousand days, allowing their proper motions to be measured individually, as shown in Fig.~\ref{fig:18hyz_X}. For more distant systems, however, or in cases where the two components are not spatially resolved, only the emission centroid of the combined image would be measurable as a point source, and the discussion below remains applicable to such cases. Emission centroids calculated by Eq.~\eqref{eq:Xcen} are shown in Fig.~\ref{fig:18hyz_im} as a red cross.

For the delayed outflow with the spherical symmetry, by definition the centroid stays at the origin. Even if the outflow has a more complicated geometry, the motion of the centroid is Newtonian and limited within an extent of $\lesssim (c\beta)T$. This limit should remain valid even for an outflow with multiple components. In such a case, different components may dominate the emission at different moments, and the emission centroid may track the brightest region and thus exhibit an apparent velocity exceeding $c\beta$. However, the centroid position itself is always confined within a distance $\lesssim (c\beta)T$ from the origin.

\begin{figure}
    \centering
    \includegraphics[width=85mm,bb=0 0 313 214]{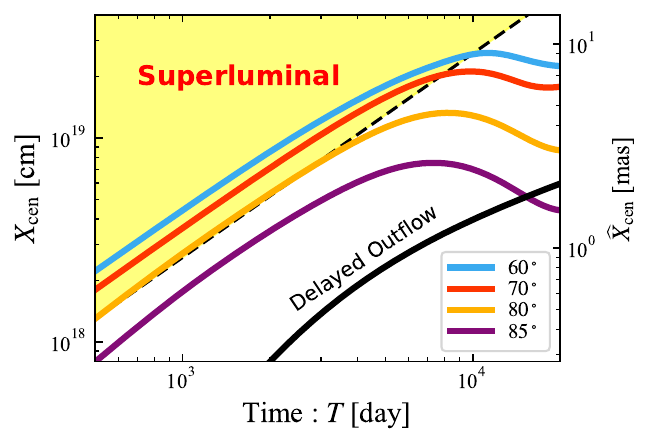}
    \caption{The evolution of the combined emission centroid by both approaching and counter jets with different viewing angles. The yellow shaded region represents the distance accessible only for a superluminal emission source ($X_{\rm cen}\geq cT$). The black curve shows the radial size of the delayed outflow.}
    \label{fig:18hyz_Xcen}
\end{figure}

For the off-axis jet, the combined centroid traces the approaching jet until the counter jet starts contributing to the emission. Then the centroid moves backward and eventually returns to the origin. Again, in the early relativistic phase the centroid exhibits superluminal motion as the approaching jet. Figure~\ref{fig:18hyz_Xcen} depicts the time evolution of the combined emission centroid by the approaching and counter jets. For a fixed jet velocity, the apparent velocity increases for a smaller viewing angle up to $\theta_{\rm obs}=1/\Gamma$ (see Eq.~\ref{eq:beta_app}). In addition, a smaller $\theta_{\rm obs}$ delays the emergence of the counter-jet (see Fig.~\ref{fig:18hyz_jet}), sustaining the superluminal motion longer. On the other hand, for larger $\theta_{\rm obs}$, the contribution from the counter-jet is no longer negligible, which makes the motion of the centroid subluminal and indistinguishable from that of a Newtonian outflow. The critical viewing angle above which $X_{\rm cen}\lesssim cT$, can be derived from Eq.~\eqref{eq:Fnu rise}. In contrast to Eq.~\eqref{eq:Fnu jet2}, without adopting the approximation of $1-\cos\theta_{\rm obs} \simeq \theta_{\rm obs}^2/2$, we obtain the angular dependence of $F_{\nu_{\rm obs}}\propto(1-\cos\theta_{\rm obs})^{p-10}=(1-\vartheta)^{p-10}$, where $\vartheta\equiv\pi/2-\theta_{\rm obs}\ll1$. Noting the radius of the jet center is given by Eq.~\eqref{eq:R}, one can obtain the location of the emission centroid by a flux-weighted mean of the position of both sided jets:
\begin{align}
X_{\rm cen}&= \frac{cT}{2} \left[\sin\theta_{\rm obs}(1-\vartheta)^{p-11}-\sin\theta_{\rm obs}(1+\vartheta)^{p-11}\right]
    \nonumber\\
&\simeq (11-p)\vartheta cT\ .
\end{align}
We approximate $\sin\theta_{\rm obs}\simeq1$ in the second equality. By requiring $X_{\rm cen}<cT$, the critical angle is given by $(11-p)\vartheta<1$, or equivalently
\begin{align}
\theta_{\rm obs} \gtrsim \frac{\pi}{2}-\frac{1}{11-p} \overset{p=2.2}{\simeq} 83^\circ\ ,
\end{align}
which roughly agrees with our result. While the apparent velocity is Newtonian for larger $\theta_{\rm obs}$, the centroid should move backward at some point, allowing distinction from the delayed Newtonian outflow.

\section{Very Long-time evolution}
\label{sec:long time}

In the previous section, we focused on AT2018hyz and its evolution up to $\sim10^4\,\rm days$ $(\simeq30\,\rm yrs)$. Over longer timescales, the jet image also exhibits several interesting behaviors. Although such a long timescale exceeds the period over which a single human researcher can work, it may be possible to observe the same phenomena in a shorter timescale for less energetic events such as other off-axis jetted TDEs with lower energy and Galactic microquasars.

\begin{figure}
    \centering
    \includegraphics[width=85mm,bb=0 0 275 185]{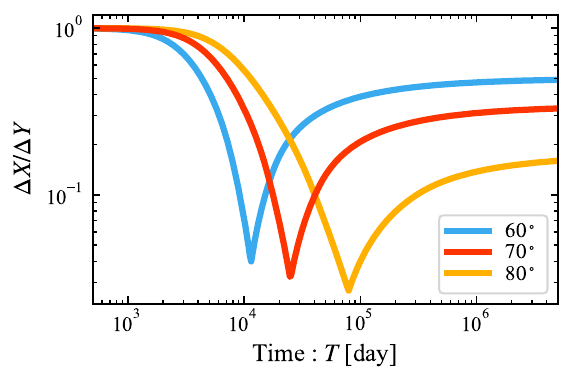}
    \caption{The long-time evolution of the aspect ratio of the approaching jet image for different viewing angles.}
    \label{fig:18hyz_ratio}
\end{figure}

The shape of the off-axis approaching jet image shows a characteristic evolution reflecting its dynamics. Figure~\ref{fig:18hyz_ratio} depicts the long-time evolution of the aspect ratio of the image defined as the ratio of the image widths in the $X$ and $Y$ directions. Initially the image is a disk with an aspect ratio of unity as seen in Fig.~\ref{fig:18hyz_im_early}, but it decreases with time, meaning that the image is elongated in the $Y$ direction (see the lower panels in Fig.~\ref{fig:18hyz_im} and Fig.~\ref{fig:18hyz_X}). Afterward, the ratio reaches a minimum at $\sim10^{4-5}\,\rm days$, then starts to increase, and eventually approaches a constant value. This behavior results from the deceleration of the jet from relativistic to Newtonian motion, which alters the geometry of the EATS.

In the early relativistic phase, the EATS has a characteristic egg-like shape clipped by the jet as discussed in the derivation of Eq.~\eqref{eq:dX}. As Fig.~\ref{fig:image_eats} illustrates, the far and near sides of the jet edge (black dots) are projected to smaller and larger $X$ coordinates. In contrast, the EATS of a spherical Newtonian outflow is simply spherical. The far and near sides are projected to larger and smaller $X$ coordinates in the image, respectively. Therefore, as the jet decelerates, the projected locations of the far and near sides of the jet edge flip on the sky plane, and the width 
$\Delta X$ has a minimum.

We derive the expressions of the aspect ratio in both relativistic and Newtonian limits. The width in the $Y$ direction is simply given by Eq.~\eqref{eq:dY} in both limits. For the relativistic regime, the expression of $\Delta X$ is already given in Eq.~\eqref{eq:dX}. In the Newtonian regime, the EATS is a sphere and a simple geometrical argument gives
\begin{align}
\Delta X = 2R_{\rm j}\theta_{\rm j} \cos\theta_{\rm obs}\ .
\end{align}
In summary, the asymptotic value of the aspect ratio is
\begin{align}
\frac{\Delta X}{\Delta Y} \simeq
\begin{cases}
1&\text{: Relativistic,}\\
\cos\theta_{\rm obs}&\text{: Newtonian.}
\end{cases}
\end{align}
Here the relativistic (Newtonian) limit corresponds to the early (late) phase of the aspect-ratio evolution in Fig.~\ref{fig:18hyz_ratio}.
The time at which the ratio takes the minimum is estimated as a conjunction point of radii of the jet center in both limits. The latter is roughly given by the energy conservation:
\begin{align}
R(T) = \left( \frac{75 E_{\rm ej}}{8\pi m_{\rm p} n} T^2 \right)^{1/5}\ .
    \label{eq:R_Sedov}
\end{align}
By equating them one can derive the timescale
\begin{align}
T_{\rm min} \simeq 2.3&\times10^{4}{\,\rm day\,}\left(\frac{n}{1\,{\rm cm^{-3}}}\right)^{-1/3}
    \nonumber\\
&\left(\frac{E_{\rm ej}}{10^{56}{\,\rm erg}}\right)^{1/3}\left(1 - \cos\theta_{\rm obs}\right)^{5/3}\ .
    \label{eq:Tmin}
\end{align}

\begin{figure}
    \centering
    \includegraphics[width=85mm,bb=0 0 276 214]{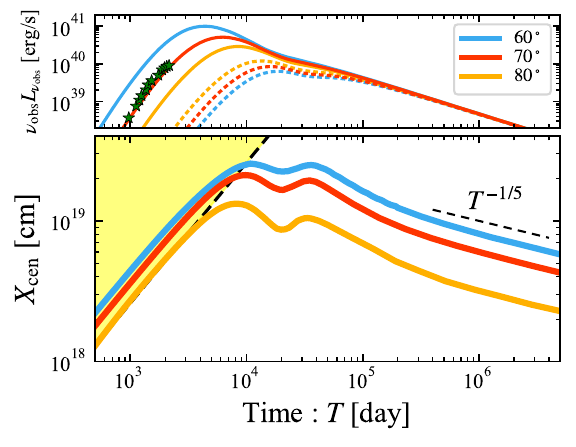}
    \caption{The long time evolution of the radio luminosity at 5 GHz (top), and the location of the emission centroid (bottom) for various viewing angles. In the top panel, the solid and dotted curves represent the luminosity of the approaching and counter jets, respectively, as in Fig.~\ref{fig:18hyz_jet}.}
    \label{fig:18hyz_Xcen_long}
\end{figure}

In addition to the image, the location of the combined emission centroid also shows a characteristic evolution. As we discussed in Sec.~\ref{sec:image}, the centroid comes back to the origin as the luminosities of both jets become comparable at late time. Counterintuitively, this occurs not instantaneously but gradually as shown in Fig.~\ref{fig:18hyz_Xcen_long}. The centroid reaches a maximum in the $X$ coordinate at $\sim 10^{4}\,\rm days$ (see also Fig.~\ref{fig:18hyz_Xcen}), exhibits a shallow dip, and then \textit{slowly} returns toward the origin. The dip corresponds to the transition into the deep Newtonian phase \citep{Huang&Cheng2003,Sironi&Giannios2013}, during which the luminosity from the approaching jet begins to decline more slowly, temporarily reducing the relative contribution from the counter jet. Once the counter jet enters the deep Newtonian phase, the two jets become nearly symmetric for $T\gtrsim10^5\,\rm days$.

Naively, one might expect the centroid to return to the origin shortly after the brightness symmetry is established. However, in practice the recession follows a power-law in 
$T$, because a small optical path difference exits. For photons to arrive at an observer at $T$, they should be emitted from the approaching and counter jets at 
\begin{align}
R_{\rm j,\substack{\rm ap\\\rm co}}=\frac{c\beta T}{1\mp\beta\cos\theta_{\rm obs}}\ ,
\end{align}
where the subscripts ``ap'' and ``co'' denote the approaching and counter jets, respectively. It should be noted that different from Eq.~\eqref{eq:R}, the velocity $\beta$ is necessary in the Newtonian limit. In short, the photons from the counter jet should be emitted earlier to compensate the traveling time for the path difference and be received at $T$. This means the counter jet is closer to the origin than the approaching jet. The location of the centroid is given by
\begin{align}
X_{\rm cen} \simeq \sin\theta_{\rm obs}R_{\rm j,ap}-\sin\theta_{\rm obs}R_{\rm j,co} \propto \beta^2 T\ ,
\end{align}
where $\beta\cos\theta_{\rm obs}\ll1$. Although the above argument assumes a constant velocity and equal luminosities from both jets, the scaling is generally valid in the Newtonian regime. In the deceleration phase, $\beta\propto T^{-3/5}$, yielding $X_{\rm cen} \propto T^{-1/5}$ consistent with our result (Fig.~\ref{fig:18hyz_Xcen_long}). That being discussed, in the case of jetted TDEs, both jets may be resolved into two distinct sources at such late time due to a large separation. Therefore, the above discussion may be meaningful for more compact sources such as microquasars.

While we have mentioned that our results apply not only to powerful off-axis jetted TDEs but also to other astrophysical transients, we have not explicitly explained why such a generalization is possible. The evolution of the jet image can be scaled through the single parameter $E_{\rm ej}/n$ (for a homogeneous CNM, and naturally generalized to other values of $k$), reflecting the well-known self-similar evolution of blast waves \citep{Taylor1950,Sedov1959,Blandford&McKee1976}. During the relativistic phase, the jet evolution is described by Eq.~\eqref{eq:R}, with $\Gamma \propto (E_{\rm ej}/nR^3)^{1/2}$, until the flow becomes Newtonian at the timescale given by Eq.~\eqref{eq:Tmin}. At later times, the evolution follows the standard Newtonian blast-wave solution, with the radius given by Eq.~\eqref{eq:R_Sedov}. Since the radio image is obtained by projecting the emitting region onto the sky plane, the entire image evolution inherits this self-similarity. Consequently, both the spatial and temporal scales of the image can be rescaled using the parameter $E_{\rm ej}/n$. Therefore, the image evolution presented in this work can be readily applied to lower-energy systems by appropriately rescaling the spatial and temporal axes. We note, however, that this scaling is valid only for moderately relativistic flows viewed sufficiently off-axis ($\theta_{\rm obs}\gg1/\Gamma$), where the one-zone approximation remains applicable.  

Finally, we note a caveat regarding the above results. We calculated the jet image at very late time under the assumption that the jet keeps its original top-hat geometry and does not experience a side-way expansion. This may be a good approximation for TDEs with powerful jets until the trans-relativistic phase, but may not be justified in the Newtonian phase. However, our findings, the evolution of the aspect ratio and the gradually receding emission centroid, could remain unchanged. In the former result, the ratio takes the minimum in the trans-relativistic phase, and for the latter one, the optical path difference still exists regardless of the jet geometry. In addition, we note recent hydrodynamical simulations find a Newtonian conical outflow expands almost keeping its original geometry \citep[e.g.,][]{Mou2025}.

\section{Summary}
\label{sec:summary}

The origin of late-time radio flares in TDEs remains uncertain. Among these events, the radio flare associated with AT2018hyz is particularly enigmatic, exhibiting an unprecedentedly sharp and sustained rise and a luminosity comparable to those of jetted TDEs \citep{Cendes+2022b,Sfaradi+2024,Cendes+2026}. The leading scenarios for the radio flare of AT2018hyz involve either an interaction between the CNM and a delayed outflow launched approximately $1000\,\rm days$ after the optical discovery, or an off-axis jet launched around the time of discovery but oriented away from our line of sight. Analyses based on the radio light curve and spectrum alone have difficulty distinguishing between these two scenarios. Radio imaging with VLBI therefore provides the most direct diagnostics to break this degeneracy \citep{Matsumoto&Piran2023}. 

In this paper, we synthesize radio images for both scenarios. In the delayed outflow scenario, the radio image directly reflects the geometry of the outflow. In our case, a spherical outflow produces a disk-like or even ring-like image due to the limb-brightening. The characteristic size is set by the radial extent of the outflow, $\sim (c\beta)T$, and increases linearly with time until significant deceleration occurs. In this geometry, the emission centroid remains stationary at the origin. Even if the outflow deviates from spherical symmetry and consists of multiple components, the centroid motion is confined within a similar range of $\lesssim (c\beta)T$.

In the off-axis jet scenario, the radio image is initially dominated by the approaching jet, and most importantly, the emission centroid exhibits superluminal motion, which would be a smoking-gun signature of a relativistic jet. As the jet decelerates, the counter-jet begins to contribute to the emission. At early times, the image of the approaching jet has a disk-like shape, similar to that of a spherical delayed outflow, but it later becomes elongated in the direction perpendicular to the jet motion. The aspect ratio of the image reaches a minimum and then increases, asymptotically approaching a constant value. This non-monotonic evolution reflects the changing geometry of the EATS. We note that these results for off-axis jets should also apply to other astrophysical systems, such as microquasars and gamma-ray bursts due to the self-similarity of the outflows.

Thus far, we have focused on radio VLBI observations, which provide the most direct way to resolve the source structure. However, if one is interested only in the location of the emission centroid rather than resolving the image, precise astrometry may also be achieved with optical or infrared (IR) telescopes, as demonstrated for GW170817 with \textit{HST} \citep{Mooley+2022}. Although the diffraction limit of $\sim100\,\rm mas$ prevents direct imaging of the source structure, the emission centroid can still be determined with high precision, provided that the source is sufficiently bright. For AT2018hyz with $p\simeq2.2$, the optical/IR luminosity may be comparable to that in the radio band, $\sim10^{40}\,\rm erg\,s^{-1}$. In the crowded nuclear region of the host galaxy, contamination from stellar light is expected to be severe in the optical band, making IR observations more suitable. If the synchrotron afterglow outshines the surrounding stellar emission, the position of the point source could be measured and potentially used to confirm superluminal motion with facilities such as \textit{JWST}.

\begin{acknowledgements}
We thank Yvette Cendes for fruitful discussions on late-time radio observations, Kazuya Takahashi for motivating the author to initiate this work, and an anonymous referee for their helpful comments. We are also grateful to Kenta Hotokezaka and Jonathan Granot for stimulating discussions. This research is supported by JSPS KAKENHI (grant No. 24K17088).

\end{acknowledgements}

\appendix
\section{Synchrotron Emission}
\label{sec:appendix}

This Appendix summarizes the basic ingredients required to compute the synchrotron emissivity and absorption coefficient. We follow the standard synchrotron emission model \citep{Sari+1998}, extending it to smoothly connect the relativistic and Newtonian regimes. Following the passage of the shock, the particle number and internal energy densities are determined by the shock jump conditions:
\begin{align}
n^\prime\simeq4\Gamma n{\,\,\,\rm and \,\,\,}e_{\rm in}^\prime\simeq m_{\rm p}c^2(\Gamma-1)n^\prime\ .
\end{align}
Assuming fractions $\varepsilon_{\rm B}$ and $\varepsilon_{\rm e}$ of the internal energy are converted to the energies of the magnetic field and non-thermal electrons, respectively, one obtains the magnetic field
\begin{align}
B^\prime\simeq(8\pi \varepsilon_{\rm B}e_{\rm in}^\prime)^{1/2}\ ,
\end{align}
and the minimum Lorentz factor of electrons
\begin{align}
\gamma_{\rm m}=\varepsilon_{\rm e}\left(\frac{p-2}{p-1}\right)\frac{m_{\rm p}}{m_{\rm e}}(\Gamma-1)\ ,
    \label{eq:gamma_m}
\end{align}
where $m_{\rm e}$ is the electron mass, and the electrons have a power-law distribution, $dn_{\rm e}/d\gamma\propto \gamma^{-p}$. For outflows with a lower velocity
\begin{align}
\beta<\beta_{\rm DN}\equiv\left[\frac{4}{\varepsilon_{\rm e}}\left(\frac{p-1}{p-2}\right)\frac{m_{\rm e}}{m_{\rm p}}\right]^{1/2}\overset{p=2.2}{\simeq}0.21\,\left(\frac{\varepsilon_{\rm e}}{0.3}\right)^{-1/2}\ ,
\end{align}
Eq.~\eqref{eq:gamma_m} yields an unphysical value, $\gamma_{\rm m}\lesssim1$. In this deep-Newtonian phase we fix $\gamma_{\rm m}=2$ and multiply $n^\prime$ in Eqs.~\eqref{eq:j_peak} and \eqref{eq:alpha} by a factor $f_{\rm DN}\equiv(\beta/\beta_{\rm DN})^2$ \citep{Matsumoto&Piran2021b}.

The synchrotron frequency and emissivity of single electron with $\gamma$ are given by 
\begin{align}
\nu^\prime(\gamma)=\frac{\gamma^2eB^\prime}{2\pi m_{\rm e}c}{\,\,\,\rm and\,\,\,}P^\prime(\gamma)=\frac{4}{3}\sigma_{\rm T}c\gamma^2\frac{(B^\prime)^2}{8\pi}\ ,
    \label{eq:nu_syn}
\end{align}
respectively, where $e$ and $\sigma_{\rm T}$ are the elementary charge and Thomson cross section, respectively. The critical Lorentz factor above which electrons are under fast cooling, is given by equating the electron's energy $\gamma_{\rm c}m_{\rm e}c^2$ and emitted energy during a dynamical timescale $P^\prime(\gamma_{\rm c})t^\prime$:
\begin{align}
\gamma_{\rm c}=\frac{6\pi m_{\rm e}c}{\sigma_{\rm T}(B^\prime)^2t^\prime}\ ,
\end{align}
where $t^\prime=t/\Gamma$ is the dynamical time in the fluid rest frame. Then the synchrotron emissivity in the slow cooling regime ($\gamma_{\rm m}<\gamma_{\rm c}$) is given by
\begin{align}
j^\prime_{\nu^\prime}=(j^\prime_{\nu^\prime})_{\rm peak}\begin{cases}
\left(\frac{\nu^\prime}{\nu^\prime_{\rm m}}\right)^{1/3}&:\nu^\prime<\nu_{\rm m}^{\prime}\ ,\\
\left(\frac{\nu^\prime}{\nu^\prime_{\rm m}}\right)^{\frac{1-p}{2}}&:\nu_{\rm m}^{\prime}<\nu^\prime<\nu^\prime_{\rm c}\ ,\\
\left(\frac{\nu_{\rm c}^\prime}{\nu^\prime_{\rm m}}\right)^{\frac{1-p}{2}}\left(\frac{\nu^\prime}{\nu^\prime_{\rm m}}\right)^{-\frac{p}{2}}&:\nu_{\rm c}^{\prime}<\nu^\prime\ ,\\
\end{cases}
\end{align}
where $\nu_{\rm m}^\prime=\nu^\prime(\gamma_{\rm m})$, $\nu_{\rm c}^\prime=\nu^\prime(\gamma_{\rm c})$, and 
\begin{align}
(j^\prime_{\nu^\prime})_{\rm peak}=\frac{n^\prime (P^\prime/\nu^\prime)}{4\pi}=\frac{\sigma_{\rm T}m_{\rm e}c^2n^\prime B^\prime}{12\pi e}\ .
    \label{eq:j_peak}
\end{align}

The absorption coefficient is given by the standard formula in \cite{Rybicki&Lightman1979} (their Eq.~6.50). For an electron population with a single power-law distribution, it can be written as
\begin{align}
\alpha^\prime_{\nu^\prime}=\frac{(p+2)(p-1)\sqrt{3}e^3n^\prime B^\prime}{16\pi m_{\rm e}^2c^2\gamma_{\rm m}(\nu^\prime)^2}\left(\frac{\nu^\prime}{\nu_{\rm m}^\prime}\right)^{-p/2} I_p(\nu^\prime/\nu_{\rm m}^\prime)\ ,
    \label{eq:alpha}
\end{align}
where 
\begin{align}
I_p(x)&\equiv\int_0^xdyy^{\frac{p-2}{2}}F(y)\\
&\simeq\begin{cases}
\frac{2^{\frac{p+2}{2}}}{p+2}\Gamma\left(\frac{3p+22}{12}\right)\Gamma\left(\frac{3p+2}{12}\right): x\gg1\ ,\\
\frac{2^{8/3}3^{1/2}\pi}{(3p+2)\Gamma(1/3)}x^{\frac{3p+2}{6}}: x\ll1\ ,\\
\end{cases}
    \nonumber\\
F(x)&\equiv x\int_x^\infty K_{5/3}(y)dy\ ,
\end{align}
where $\Gamma(x)$ and $K_{5/3}(x)$ are the Gamma function and modified Bessel function, respectively. We remark that \cite{Rybicki&Lightman1979} originally obtain the absorption coefficient (their Eq.~6.53) in the limit of $\nu\gg\nu_{\rm m}$ keeping the dependence on the electron pitch angle $\theta_{\rm p}$ and adopting a slightly different definition of the characteristic frequency, $\nu^\prime(\gamma)={3\gamma^2eB^\prime\sin\theta_{\rm p}}/{4\pi m_{\rm e}c}$, instead of our Eq.~\eqref{eq:nu_syn}. Their equation is identical to ours by introducing $\nu_{\rm m}$ and setting $\sin\theta_{\rm p}\to1$.

%#\bibliographystyle{mn2e}
\bibliographystyle{aasjournalv7}
%\bibliography{refs}
\bibliography{reference_matsumoto}

\begin{thebibliography}{}
\expandafter\ifx\csname natexlab\endcsname\relax\def\natexlab#1{#1}\fi
\providecommand{\url}[1]{\href{#1}{#1}}
\providecommand{\dodoi}[1]{doi:~\href{http://doi.org/#1}{\nolinkurl{#1}}}
\providecommand{\doeprint}[1]{\href{http://ascl.net/#1}{\nolinkurl{http://ascl.net/#1}}}
\providecommand{\doarXiv}[1]{\href{https://arxiv.org/abs/#1}{\nolinkurl{https://arxiv.org/abs/#1}}}

% type= article
\bibitem[{K.~D. {Alexander} {et~al.}(2026){Alexander}, {Margutti}, {Gomez},
  {Stroh}, {Chornock}, {Laskar}, {Cendes}, {Berger}, {Eftekhari}, {Franz},
  {Hajela}, {Metzger}, {Terreran}, {Bietenholz}, {Christy}, {De Colle},
  {Komossa}, {Nicholl}, {Ramirez-Ruiz}, {Saxton}, {Schroeder}, {Williams}, \&
  {Wu}}]{Alexander+2026}
{Alexander}, K.~D., {Margutti}, R., {Gomez}, S., {et~al.} 2026,
  \bibinfo{title}{{The Multiwavelength Context of Delayed Radio Emission in
  Tidal Disruption Events: Evidence for Accretion-driven Outflows},} \apj,
  1000, 139, \dodoi{10.3847/1538-4357/ae40ab}

% type= article
\bibitem[{R. {Barniol Duran} {et~al.}(2013){Barniol Duran}, {Nakar}, \&
  {Piran}}]{BarniolDuran+2013}
{Barniol Duran}, R., {Nakar}, E., \& {Piran}, T. 2013, \bibinfo{title}{{Radius
  Constraints and Minimal Equipartition Energy of Relativistically Moving
  Synchrotron Sources},} \apj, 772, 78, \dodoi{10.1088/0004-637X/772/1/78}

% type= article
\bibitem[{P. {Beniamini} {et~al.}(2023){Beniamini}, {Piran}, \&
  {Matsumoto}}]{Beniamini+2023b}
{Beniamini}, P., {Piran}, T., \& {Matsumoto}, T. 2023, \bibinfo{title}{{Swift
  J1644+57 as an off-axis Jet},} \mnras, 524, 1386,
  \dodoi{10.1093/mnras/stad1950}

% type= article
\bibitem[{E. {Berger} {et~al.}(2012){Berger}, {Zauderer}, {Pooley},
  {Soderberg}, {Sari}, {Brunthaler}, \& {Bietenholz}}]{Berger+2012}
{Berger}, E., {Zauderer}, A., {Pooley}, G.~G., {et~al.} 2012,
  \bibinfo{title}{{Radio Monitoring of the Tidal Disruption Event Swift
  J164449.3+573451. I. Jet Energetics and the Pristine Parsec-scale Environment
  of a Supermassive Black Hole},} \apj, 748, 36,
  \dodoi{10.1088/0004-637X/748/1/36}

% type= article
\bibitem[{R.~D. {Blandford} \& C.~F. {McKee}(1976){Blandford} \&
  {McKee}}]{Blandford&McKee1976}
{Blandford}, R.~D., \& {McKee}, C.~F. 1976, \bibinfo{title}{{Fluid dynamics of
  relativistic blast waves},} Physics of Fluids, 19, 1130,
  \dodoi{10.1063/1.861619}

% type= article
\bibitem[{G. {Bruni} {et~al.}(2021){Bruni}, {O'Connor}, {Matsumoto}, {Troja},
  {Piran}, {Piro}, \& {Ricci}}]{Bruni+2021}
{Bruni}, G., {O'Connor}, B., {Matsumoto}, T., {et~al.} 2021,
  \bibinfo{title}{{Late-time radio observations of the short GRB 200522A:
  constraints on the magnetar model},} \mnras, 505, L41,
  \dodoi{10.1093/mnrasl/slab046}

% type= article
\bibitem[{Y. {Cendes} {et~al.}(2021){Cendes}, {Eftekhari}, {Berger}, \&
  {Polisensky}}]{Cendes+2021}
{Cendes}, Y., {Eftekhari}, T., {Berger}, E., \& {Polisensky}, E. 2021,
  \bibinfo{title}{{Radio Monitoring of the Tidal Disruption Event Swift
  J164449.3+573451. IV. Continued Fading and Non-relativistic Expansion},}
  \apj, 908, 125, \dodoi{10.3847/1538-4357/abd323}

% type= article
\bibitem[{Y. {Cendes} {et~al.}(2022){Cendes}, {Berger}, {Alexander}, {Gomez},
  {Hajela}, {Chornock}, {Laskar}, {Margutti}, {Metzger}, {Bietenholz},
  {Brethauer}, \& {Wieringa}}]{Cendes+2022b}
{Cendes}, Y., {Berger}, E., {Alexander}, K.~D., {et~al.} 2022,
  \bibinfo{title}{{A Mildly Relativistic Outflow Launched Two Years after
  Disruption in Tidal Disruption Event AT2018hyz},} \apj, 938, 28,
  \dodoi{10.3847/1538-4357/ac88d0}

% type= article
\bibitem[{Y. {Cendes} {et~al.}(2024){Cendes}, {Berger}, {Alexander},
  {Chornock}, {Margutti}, {Metzger}, {Wieringa}, {Bietenholz}, {Hajela},
  {Laskar}, {Stroh}, \& {Terreran}}]{Cendes+2024}
{Cendes}, Y., {Berger}, E., {Alexander}, K.~D., {et~al.} 2024,
  \bibinfo{title}{{Ubiquitous Late Radio Emission from Tidal Disruption
  Events},} \apj, 971, 185, \dodoi{10.3847/1538-4357/ad5541}

% type= article
\bibitem[{Y. {Cendes} {et~al.}(2026){Cendes}, {Berger}, {Beniamini}, {Gill},
  {Matsumoto}, {Alexander}, {Bietenholz}, {Hajela}, {Christy}, {Chornock},
  {Gomez}, {Gurwell}, {Keating}, {Laskar}, {Margutti}, {Rao}, {Velez}, \&
  {Wieringa}}]{Cendes+2026}
{Cendes}, Y., {Berger}, E., {Beniamini}, P., {et~al.} 2026,
  \bibinfo{title}{{Continued Rapid Radio Brightening of the Tidal Disruption
  Event AT2018hyz},} \apj, 998, 111, \dodoi{10.3847/1538-4357/ae286d}

% type= article
\bibitem[{C.~T. {Christy} {et~al.}(2024){Christy}, {Alexander}, {Margutti},
  {Wieringa}, {Cendes}, {Chornock}, {Laskar}, {Berger}, {Bietenholz},
  {Coppejans}, {De Colle}, {Eftekhari}, {Holoien}, {Matsumoto}, {Miller-Jones},
  {Ramirez-Ruiz}, {Saxton}, \& {van Velzen}}]{Christy+2024}
{Christy}, C.~T., {Alexander}, K.~D., {Margutti}, R., {et~al.} 2024,
  \bibinfo{title}{{The Peculiar Radio Evolution of the Tidal Disruption Event
  ASASSN-19bt},} \apj, 974, 18, \dodoi{10.3847/1538-4357/ad675b}

% type= article
\bibitem[{T. {Eftekhari} {et~al.}(2018){Eftekhari}, {Berger}, {Zauderer},
  {Margutti}, \& {Alexander}}]{Eftekhari+2018}
{Eftekhari}, T., {Berger}, E., {Zauderer}, B.~A., {Margutti}, R., \&
  {Alexander}, K.~D. 2018, \bibinfo{title}{{Radio Monitoring of the Tidal
  Disruption Event Swift J164449.3+573451. III. Late-time Jet Energetics and a
  Deviation from Equipartition},} \apj, 854, 86,
  \dodoi{10.3847/1538-4357/aaa8e0}

% type= incollection
\bibitem[{R. {Fender}(2006){Fender}}]{Fender2006}
{Fender}, R. 2006, \bibinfo{title}{{Jets from X-ray binaries},} in Compact
  stellar X-ray sources, ed. W.~H.~G. {Lewin} \& M.~{van der Klis}, Vol.~39,
  381--419, \dodoi{10.48550/arXiv.astro-ph/0303339}

% type= article
\bibitem[{J.~J. {Fern{\'a}ndez} {et~al.}(2022){Fern{\'a}ndez}, {Kobayashi}, \&
  {Lamb}}]{Fernandez+2022}
{Fern{\'a}ndez}, J.~J., {Kobayashi}, S., \& {Lamb}, G.~P. 2022,
  \bibinfo{title}{{Lateral spreading effects on VLBI radio images of neutron
  star merger jets},} \mnras, 509, 395, \dodoi{10.1093/mnras/stab2879}

% type= article
\bibitem[{G. {Ghirlanda} {et~al.}(2019){Ghirlanda}, {Salafia}, {Paragi},
  {Giroletti}, {Yang}, {Marcote}, {Blanchard}, {Agudo}, {An}, {Bernardini},
  {Beswick}, {Branchesi}, {Campana}, {Casadio}, {Chassand e-Mottin}, {Colpi},
  {Covino}, {D'Avanzo}, {D'Elia}, {Frey}, {Gawronski}, {Ghisellini}, {Gurvits},
  {Jonker}, {van Langevelde}, {Melandri}, {Moldon}, {Nava}, {Perego},
  {Perez-Torres}, {Reynolds}, {Salvaterra}, {Tagliaferri}, {Venturi},
  {Vergani}, \& {Zhang}}]{Ghirlanda+2019}
{Ghirlanda}, G., {Salafia}, O.~S., {Paragi}, Z., {et~al.} 2019,
  \bibinfo{title}{{Compact radio emission indicates a structured jet was
  produced by a binary neutron star merger},} Science, 363, 968,
  \dodoi{10.1126/science.aau8815}

% type= article
\bibitem[{D. {Giannios} \& B.~D. {Metzger}(2011){Giannios} \&
  {Metzger}}]{Giannios&Metzger2011}
{Giannios}, D., \& {Metzger}, B.~D. 2011, \bibinfo{title}{{Radio transients
  from stellar tidal disruption by massive black holes},} \mnras, 416, 2102,
  \dodoi{10.1111/j.1365-2966.2011.19188.x}

% type= article
\bibitem[{R. {Gill} \& J. {Granot}(2018){Gill} \& {Granot}}]{Gill&Granot2018b}
{Gill}, R., \& {Granot}, J. 2018, \bibinfo{title}{{Afterglow imaging and
  polarization of misaligned structured GRB jets and cocoons: breaking the
  degeneracy in GRB 170817A},} \mnras, 478, 4128, \dodoi{10.1093/mnras/sty1214}

% type= article
\bibitem[{W.~W. {Golay} {et~al.}(2025){Golay}, {Berger}, {Cendes}, {Masterson},
  {Polisensky}, {Mutel}, {Blanchard}, {Kumar}, {Margutti}, {Drout},
  {Panagiotou}, {De}, \& {Kara}}]{Golay+2025}
{Golay}, W.~W., {Berger}, E., {Cendes}, Y., {et~al.} 2025,
  \bibinfo{title}{{Radio Emission from the Infrared Tidal Disruption Event
  WTP14adeqka: The First Directly Resolved Delayed Outflow from a TDE},} arXiv
  e-prints, arXiv:2508.16756, \dodoi{10.48550/arXiv.2508.16756}

% type= article
\bibitem[{S. {Gomez} {et~al.}(2020){Gomez}, {Nicholl}, {Short}, {Margutti},
  {Alexander}, {Blanchard}, {Berger}, {Eftekhari}, {Schulze}, {Anderson},
  {Arcavi}, {Chornock}, {Cowperthwaite}, {Galbany}, {Herzog}, {Hiramatsu},
  {Hosseinzadeh}, {Laskar}, {M{\"u}ller Bravo}, {Patton}, \&
  {Terreran}}]{Gomez+2020}
{Gomez}, S., {Nicholl}, M., {Short}, P., {et~al.} 2020, \bibinfo{title}{{The
  Tidal Disruption Event AT 2018hyz II: Light-curve modelling of a partially
  disrupted star},} \mnras, 497, 1925, \dodoi{10.1093/mnras/staa2099}

% type= article
\bibitem[{A.~J. {Goodwin} {et~al.}(2023){Goodwin}, {Alexander}, {Miller-Jones},
  {Bietenholz}, {van Velzen}, {Anderson}, {Berger}, {Cendes}, {Chornock},
  {Coppejans}, {Eftekhari}, {Gezari}, {Laskar}, {Ramirez-Ruiz}, \&
  {Saxton}}]{Goodwin+2023b}
{Goodwin}, A.~J., {Alexander}, K.~D., {Miller-Jones}, J.~C.~A., {et~al.} 2023,
  \bibinfo{title}{{A radio-emitting outflow produced by the tidal disruption
  event AT2020vwl},} \mnras, 522, 5084, \dodoi{10.1093/mnras/stad1258}

% type= article
\bibitem[{A.~J. {Goodwin} {et~al.}(2025){Goodwin}, {Mummery}, {Laskar},
  {Alexander}, {Anderson}, {Bietenholz}, {Bonnerot}, {Christy}, {Golay}, {Lu},
  {Margutti}, {Miller-Jones}, {Ramirez-Ruiz}, {Saxton}, \& {van
  Velzen}}]{Goodwin+2025}
{Goodwin}, A.~J., {Mummery}, A., {Laskar}, T., {et~al.} 2025,
  \bibinfo{title}{{A Second Radio Flare from the Tidal Disruption Event
  AT2020vwl: A Delayed Outflow Ejection?},} \apj, 981, 122,
  \dodoi{10.3847/1538-4357/adb0b1}

% type= article
\bibitem[{T. {Govreen-Segal} \& E. {Nakar}(2023){Govreen-Segal} \&
  {Nakar}}]{GovreenSegal&Nakar2023}
{Govreen-Segal}, T., \& {Nakar}, E. 2023, \bibinfo{title}{{Analytic model for
  off-axis GRB afterglow images - geometry measurement and implications for
  measuring H$_{0}$},} \mnras, 524, 403, \dodoi{10.1093/mnras/stad1628}

% type= article
\bibitem[{J. {Granot} {et~al.}(2018){Granot}, {De Colle}, \&
  {Ramirez-Ruiz}}]{Granot+2018}
{Granot}, J., {De Colle}, F., \& {Ramirez-Ruiz}, E. 2018,
  \bibinfo{title}{{Off-axis afterglow light curves and images from 2D
  hydrodynamic simulations of double-sided GRB jets in a stratified external
  medium},} \mnras, 481, 2711, \dodoi{10.1093/mnras/sty2454}

% type= article
\bibitem[{J. {Granot} \& T. {Piran}(2012){Granot} \&
  {Piran}}]{Granot&Piran2012}
{Granot}, J., \& {Piran}, T. 2012, \bibinfo{title}{{On the lateral expansion of
  gamma-ray burst jets},} \mnras, 421, 570,
  \dodoi{10.1111/j.1365-2966.2011.20335.x}

% type= article
\bibitem[{J. {Granot} {et~al.}(1999{\natexlab{a}}){Granot}, {Piran}, \&
  {Sari}}]{Granot+1999b}
{Granot}, J., {Piran}, T., \& {Sari}, R. 1999{\natexlab{a}},
  \bibinfo{title}{{Synchrotron Self-Absorption in Gamma-Ray Burst Afterglow},}
  \apj, 527, 236, \dodoi{10.1086/308052}

% type= article
\bibitem[{J. {Granot} {et~al.}(1999{\natexlab{b}}){Granot}, {Piran}, \&
  {Sari}}]{Granot+1999}
{Granot}, J., {Piran}, T., \& {Sari}, R. 1999{\natexlab{b}},
  \bibinfo{title}{{Images and Spectra from the Interior of a Relativistic
  Fireball},} \apj, 513, 679, \dodoi{10.1086/306884}

% type= article
\bibitem[{A. {Hajela} {et~al.}(2025){Hajela}, {Alexander}, {Margutti},
  {Chornock}, {Bietenholz}, {Christy}, {Stroh}, {Terreran}, {Saxton},
  {Komossa}, {Bright}, {Ramirez-Ruiz}, {Coppejans}, {Leung}, {Cendes},
  {Wiston}, {Laskar}, {Horesh}, {Schroeder}, {A.~J.}, {Wieringa}, {Velez},
  {Berger}, {Blanchard}, {Eftekhari}, {Gomez}, {Nicholl}, {Sears}, \&
  {Zauderer}}]{Hajela+2025}
{Hajela}, A., {Alexander}, K.~D., {Margutti}, R., {et~al.} 2025,
  \bibinfo{title}{{Eight Years of Light from ASASSN-15oi: Toward Understanding
  the Late-time Evolution of TDEs},} \apj, 983, 29,
  \dodoi{10.3847/1538-4357/adb620}

% type= article
\bibitem[{E. {Hammerstein} {et~al.}(2023){Hammerstein}, {van Velzen}, {Gezari},
  {Cenko}, {Yao}, {Ward}, {Frederick}, {Villanueva}, {Somalwar}, {Graham},
  {Kulkarni}, {Stern}, {Andreoni}, {Bellm}, {Dekany}, {Dhawan}, {Drake},
  {Fremling}, {Gatkine}, {Groom}, {Ho}, {Kasliwal}, {Karambelkar}, {Kool},
  {Masci}, {Medford}, {Perley}, {Purdum}, {Roestel}, {Sharma}, {Sollerman},
  {Taggart}, \& {Yan}}]{Hammerstein+2023}
{Hammerstein}, E., {van Velzen}, S., {Gezari}, S., {et~al.} 2023,
  \bibinfo{title}{{The Final Season Reimagined: 30 Tidal Disruption Events from
  the ZTF-I Survey},} \apj, 942, 9, \dodoi{10.3847/1538-4357/aca283}

% type= article
\bibitem[{A.~Y.~Q. {Ho} {et~al.}(2025){Ho}, {Yao}, {Matsumoto}, {Schroeder},
  {Coughlin}, {Perley}, {Andreoni}, {Bellm}, {Chen}, {Chornock}, {Covarrubias},
  {Das}, {Fremling}, {Gilfanov}, {Hinds}, {Jarvis}, {Kasliwal}, {Liu}, {Lyman},
  {Masci}, {Prince}, {Ravi}, {Rich}, {Riddle}, {Sevilla}, {Smith}, {Sollerman},
  {Somalwar}, {Srinivasaragavan}, {Sunyaev}, {Vail}, {Wise}, \&
  {Yun}}]{Ho+2025}
{Ho}, A. Y.~Q., {Yao}, Y., {Matsumoto}, T., {et~al.} 2025, \bibinfo{title}{{A
  Luminous Red Optical Flare and Hard X-Ray Emission in the Tidal Disruption
  Event AT 2024kmq},} \apj, 989, 54, \dodoi{10.3847/1538-4357/ade8f2}

% type= article
\bibitem[{A. {Horesh} {et~al.}(2021{\natexlab{a}}){Horesh}, {Cenko}, \&
  {Arcavi}}]{Horesh+2021}
{Horesh}, A., {Cenko}, S.~B., \& {Arcavi}, I. 2021{\natexlab{a}},
  \bibinfo{title}{{Delayed radio flares from a tidal disruption event},} Nature
  Astronomy, 5, 491, \dodoi{10.1038/s41550-021-01300-8}

% type= article
\bibitem[{A. {Horesh} {et~al.}(2021{\natexlab{b}}){Horesh}, {Sfaradi},
  {Fender}, {Green}, {Williams}, \& {Bright}}]{Horesh+2021b}
{Horesh}, A., {Sfaradi}, I., {Fender}, R., {et~al.} 2021{\natexlab{b}},
  \bibinfo{title}{{Are Delayed Radio Flares Common in Tidal Disruption Events?
  The Case of the TDE iPTF 16fnl},} \apjl, 920, L5,
  \dodoi{10.3847/2041-8213/ac25fe}

% type= article
\bibitem[{F.~F. {Hu} {et~al.}(2025){Hu}, {Goodwin}, {Price}, {Mandel}, {Sari},
  \& {Hayasaki}}]{Hu+2025}
{Hu}, F.~F., {Goodwin}, A., {Price}, D.~J., {et~al.} 2025,
  \bibinfo{title}{{Radio Emission from Tidal Disruption Events Produced by the
  Collision between Super-Eddington Outflows and the Circumnuclear Medium},}
  \apjl, 988, L24, \dodoi{10.3847/2041-8213/adeb79}

% type= article
\bibitem[{Y.~F. {Huang} \& K.~S. {Cheng}(2003){Huang} \&
  {Cheng}}]{Huang&Cheng2003}
{Huang}, Y.~F., \& {Cheng}, K.~S. 2003, \bibinfo{title}{{Gamma-ray bursts:
  optical afterglows in the deep Newtonian phase},} \mnras, 341, 263,
  \dodoi{10.1046/j.1365-8711.2003.06430.x}

% type= article
\bibitem[{W. {Lu} {et~al.}(2024){Lu}, {Matsumoto}, \& {Matzner}}]{Lu+2024}
{Lu}, W., {Matsumoto}, T., \& {Matzner}, C.~D. 2024,
  \bibinfo{title}{{Misaligned precessing jets are choked by the accretion disc
  wind},} \mnras, 533, 979, \dodoi{10.1093/mnras/stae1770}

% type= article
\bibitem[{T. {Matsumoto} \& B.~D. {Metzger}(2023){Matsumoto} \&
  {Metzger}}]{Matsumoto&Metzger2023}
{Matsumoto}, T., \& {Metzger}, B.~D. 2023, \bibinfo{title}{{Synchrotron
  afterglow model for AT 2022cmc: jetted tidal disruption event or
  engine-powered supernova?},} \mnras, 522, 4028,
  \dodoi{10.1093/mnras/stad1182}

% type= article
\bibitem[{T. {Matsumoto} \& T. {Piran}(2021){Matsumoto} \&
  {Piran}}]{Matsumoto&Piran2021b}
{Matsumoto}, T., \& {Piran}, T. 2021, \bibinfo{title}{{Radio constraint on
  outflows from tidal disruption events},} \mnras, 507, 4196,
  \dodoi{10.1093/mnras/stab2418}

% type= article
\bibitem[{T. {Matsumoto} \& T. {Piran}(2023){Matsumoto} \&
  {Piran}}]{Matsumoto&Piran2023}
{Matsumoto}, T., \& {Piran}, T. 2023, \bibinfo{title}{{Generalized
  equipartition method from an arbitrary viewing angle},} \mnras, 522, 4565,
  \dodoi{10.1093/mnras/stad1269}

% type= article
\bibitem[{T. {Matsumoto} \& T. {Piran}(2024){Matsumoto} \&
  {Piran}}]{Matsumoto&Piran2024}
{Matsumoto}, T., \& {Piran}, T. 2024, \bibinfo{title}{{Late-time Radio Flares
  in Tidal Disruption Events},} \apj, 971, 49, \dodoi{10.3847/1538-4357/ad58ba}

% type= article
\bibitem[{T. {Matsumoto} {et~al.}(2022){Matsumoto}, {Piran}, \&
  {Krolik}}]{Matsumoto+2022}
{Matsumoto}, T., {Piran}, T., \& {Krolik}, J.~H. 2022, \bibinfo{title}{{What
  powers the radio emission in TDE AT2019dsg: A long-lived jet or the
  disruption itself?},} \mnras, 511, 5085, \dodoi{10.1093/mnras/stac382}

% type= article
\bibitem[{P. {Mimica} {et~al.}(2015){Mimica}, {Giannios}, {Metzger}, \&
  {Aloy}}]{Mimica+2015}
{Mimica}, P., {Giannios}, D., {Metzger}, B.~D., \& {Aloy}, M.~A. 2015,
  \bibinfo{title}{{The radio afterglow of Swift J1644+57 reveals a powerful jet
  with fast core and slow sheath},} \mnras, 450, 2824,
  \dodoi{10.1093/mnras/stv825}

% type= article
\bibitem[{I.~F. {Mirabel} \& L.~F. {Rodr{\'\i}guez}(1994){Mirabel} \&
  {Rodr{\'\i}guez}}]{Mirabel&Rodriguez1994}
{Mirabel}, I.~F., \& {Rodr{\'\i}guez}, L.~F. 1994, \bibinfo{title}{{A
  superluminal source in the Galaxy},} \nat, 371, 46, \dodoi{10.1038/371046a0}

% type= article
\bibitem[{K.~P. {Mooley} {et~al.}(2022){Mooley}, {Anderson}, \&
  {Lu}}]{Mooley+2022}
{Mooley}, K.~P., {Anderson}, J., \& {Lu}, W. 2022, \bibinfo{title}{{Optical
  superluminal motion measurement in the neutron-star merger GW170817},} \nat,
  610, 273, \dodoi{10.1038/s41586-022-05145-7}

% type= article
\bibitem[{K.~P. {Mooley} {et~al.}(2018){Mooley}, {Deller}, {Gottlieb}, {Nakar},
  {Hallinan}, {Bourke}, {Frail}, {Horesh}, {Corsi}, \&
  {Hotokezaka}}]{Mooley+2018b}
{Mooley}, K.~P., {Deller}, A.~T., {Gottlieb}, O., {et~al.} 2018,
  \bibinfo{title}{{Superluminal motion of a relativistic jet in the
  neutron-star merger GW170817},} \nat, 561, 355,
  \dodoi{10.1038/s41586-018-0486-3}

% type= article
\bibitem[{G. {Mou}(2025){Mou}}]{Mou2025}
{Mou}, G. 2025, \bibinfo{title}{{Numerical Studies on the Radio Afterglows in
  TDE (I): Forward Shock},} arXiv e-prints, arXiv:2510.14715,
  \dodoi{10.48550/arXiv.2510.14715}

% type= article
\bibitem[{G. {Mou} \& X. {Shu}(2025){Mou} \& {Shu}}]{Mou&Shu2025}
{Mou}, G., \& {Shu}, X. 2025, \bibinfo{title}{{Numerical Studies on the Radio
  Afterglows in TDE: Bow Shock},} arXiv e-prints, arXiv:2510.25033,
  \dodoi{10.48550/arXiv.2510.25033}

% type= article
\bibitem[{V. {Nedora} {et~al.}(2023){Nedora}, {Dietrich}, \&
  {Shibata}}]{Nedora+2023}
{Nedora}, V., {Dietrich}, T., \& {Shibata}, M. 2023, \bibinfo{title}{{Synthetic
  radio images of structured GRB and kilonova afterglows},} \mnras, 524, 5514,
  \dodoi{10.1093/mnras/stad2128}

% type= article
\bibitem[{T. {Piran} {et~al.}(2013){Piran}, {Nakar}, \& {Rosswog}}]{Piran+2013}
{Piran}, T., {Nakar}, E., \& {Rosswog}, S. 2013, \bibinfo{title}{{The
  electromagnetic signals of compact binary mergers},} \mnras, 430, 2121,
  \dodoi{10.1093/mnras/stt037}

% type= article
\bibitem[{ {Planck Collaboration} {et~al.}(2020){Planck Collaboration},
  {Aghanim}, {Akrami}, {Ashdown}, {Aumont}, {Baccigalupi}, {Ballardini},
  {Banday}, {Barreiro}, {Bartolo}, {Basak}, {Battye}, {Benabed}, {Bernard},
  {Bersanelli}, {Bielewicz}, {Bock}, {Bond}, {Borrill}, {Bouchet}, {Boulanger},
  {Bucher}, {Burigana}, {Butler}, {Calabrese}, {Cardoso}, {Carron},
  {Challinor}, {Chiang}, {Chluba}, {Colombo}, {Combet}, {Contreras}, {Crill},
  {Cuttaia}, {de Bernardis}, {de Zotti}, {Delabrouille}, {Delouis}, {Di
  Valentino}, {Diego}, {Dor{\'e}}, {Douspis}, {Ducout}, {Dupac}, {Dusini},
  {Efstathiou}, {Elsner}, {En{\ss}lin}, {Eriksen}, {Fantaye}, {Farhang},
  {Fergusson}, {Fernandez-Cobos}, {Finelli}, {Forastieri}, {Frailis},
  {Fraisse}, {Franceschi}, {Frolov}, {Galeotta}, {Galli}, {Ganga},
  {G{\'e}nova-Santos}, {Gerbino}, {Ghosh}, {Gonz{\'a}lez-Nuevo}, {G{\'o}rski},
  {Gratton}, {Gruppuso}, {Gudmundsson}, {Hamann}, {Handley}, {Hansen},
  {Herranz}, {Hildebrandt}, {Hivon}, {Huang}, {Jaffe}, {Jones}, {Karakci},
  {Keih{\"a}nen}, {Keskitalo}, {Kiiveri}, {Kim}, {Kisner}, {Knox},
  {Krachmalnicoff}, {Kunz}, {Kurki-Suonio}, {Lagache}, {Lamarre}, {Lasenby},
  {Lattanzi}, {Lawrence}, {Le Jeune}, {Lemos}, {Lesgourgues}, {Levrier},
  {Lewis}, {Liguori}, {Lilje}, {Lilley}, {Lindholm}, {L{\'o}pez-Caniego},
  {Lubin}, {Ma}, {Mac{\'\i}as-P{\'e}rez}, {Maggio}, {Maino}, {Mandolesi},
  {Mangilli}, {Marcos-Caballero}, {Maris}, {Martin}, {Martinelli},
  {Mart{\'\i}nez-Gonz{\'a}lez}, {Matarrese}, {Mauri}, {McEwen}, {Meinhold},
  {Melchiorri}, {Mennella}, {Migliaccio}, {Millea}, {Mitra},
  {Miville-Desch{\^e}nes}, {Molinari}, {Montier}, {Morgante}, {Moss}, {Natoli},
  {N{\o}rgaard-Nielsen}, {Pagano}, {Paoletti}, {Partridge}, {Patanchon},
  {Peiris}, {Perrotta}, {Pettorino}, {Piacentini}, {Polastri}, {Polenta},
  {Puget}, {Rachen}, {Reinecke}, {Remazeilles}, {Renzi}, {Rocha}, {Rosset},
  {Roudier}, {Rubi{\~n}o-Mart{\'\i}n}, {Ruiz-Granados}, {Salvati}, {Sandri},
  {Savelainen}, {Scott}, {Shellard}, {Sirignano}, {Sirri}, {Spencer},
  {Sunyaev}, {Suur-Uski}, {Tauber}, {Tavagnacco}, {Tenti}, {Toffolatti},
  {Tomasi}, {Trombetti}, {Valenziano}, {Valiviita}, {Van Tent}, {Vibert},
  {Vielva}, {Villa}, {Vittorio}, {Wandelt}, {Wehus}, {White}, {White},
  {Zacchei}, \& {Zonca}}]{PlanckCollaboration2020_parameter}
{Planck Collaboration}, {Aghanim}, N., {Akrami}, Y., {et~al.} 2020,
  \bibinfo{title}{{Planck 2018 results. VI. Cosmological parameters},} \aap,
  641, A6, \dodoi{10.1051/0004-6361/201833910}

% type= article
\bibitem[{M.~J. {Rees}(1966){Rees}}]{Rees1966}
{Rees}, M.~J. 1966, \bibinfo{title}{{Appearance of Relativistically Expanding
  Radio Sources},} \nat, 211, 468, \dodoi{10.1038/211468a0}

% type= article
\bibitem[{M.~J. {Rees}(1988){Rees}}]{Rees1988}
{Rees}, M.~J. 1988, \bibinfo{title}{{Tidal disruption of stars by black holes
  of {}10$^{6}$-{}10$^{8}$ solar masses in nearby galaxies},} \nat, 333, 523,
  \dodoi{10.1038/333523a0}

% type= article
\bibitem[{J.~E. {Rhoads}(1999){Rhoads}}]{Rhoads1999}
{Rhoads}, J.~E. 1999, \bibinfo{title}{{The Dynamics and Light Curves of Beamed
  Gamma-Ray Burst Afterglows},} \apj, 525, 737, \dodoi{10.1086/307907}

% type= article
\bibitem[{R. {Ricci} {et~al.}(2021){Ricci}, {Troja}, {Bruni}, {Matsumoto},
  {Piro}, {O'Connor}, {Piran}, {Navaieelavasani}, {Corsi}, {Giacomazzo}, \&
  {Wieringa}}]{Ricci+2021}
{Ricci}, R., {Troja}, E., {Bruni}, G., {et~al.} 2021,
  \bibinfo{title}{{Searching for the radio remnants of short-duration gamma-ray
  bursts},} \mnras, 500, 1708, \dodoi{10.1093/mnras/staa3241}

% type= book
\bibitem[{G.~B. {Rybicki} \& A.~P. {Lightman}(1979){Rybicki} \&
  {Lightman}}]{Rybicki&Lightman1979}
{Rybicki}, G.~B., \& {Lightman}, A.~P. 1979, {Radiative processes in
  astrophysics}

% type= article
\bibitem[{G. {Sadeh} {et~al.}(2024){Sadeh}, {Linder}, \& {Waxman}}]{Sadeh+2024}
{Sadeh}, G., {Linder}, N., \& {Waxman}, E. 2024, \bibinfo{title}{{Non-thermal
  emission from mildly relativistic dynamical ejecta of neutron star mergers:
  spectrum and sky image},} \mnras, 531, 3279, \dodoi{10.1093/mnras/stae1286}

% type= article
\bibitem[{R. {Sari}(1998){Sari}}]{Sari1998}
{Sari}, R. 1998, \bibinfo{title}{{The Observed Size and Shape of Gamma-Ray
  Burst Afterglow},} \apjl, 494, L49, \dodoi{10.1086/311160}

% type= article
\bibitem[{R. {Sari} {et~al.}(1998){Sari}, {Piran}, \& {Narayan}}]{Sari+1998}
{Sari}, R., {Piran}, T., \& {Narayan}, R. 1998, \bibinfo{title}{{Spectra and
  Light Curves of Gamma-Ray Burst Afterglows},} \apjl, 497, L17,
  \dodoi{10.1086/311269}

% type= article
\bibitem[{Y. {Sato} {et~al.}(2024){Sato}, {Murase}, {Bhattacharya}, {Carpio},
  {Mukhopadhyay}, \& {Zhang}}]{Sato+2024}
{Sato}, Y., {Murase}, K., {Bhattacharya}, M., {et~al.} 2024,
  \bibinfo{title}{{Two-component off-axis jet model for radio flares of tidal
  disruption events},} \prd, 110, L061307, \dodoi{10.1103/PhysRevD.110.L061307}

% type= book
\bibitem[{L.~I. {Sedov}(1959){Sedov}}]{Sedov1959}
{Sedov}, L.~I. 1959, {Similarity and Dimensional Methods in Mechanics}

% type= article
\bibitem[{I. {Sfaradi} {et~al.}(2024){Sfaradi}, {Beniamini}, {Horesh}, {Piran},
  {Bright}, {Rhodes}, {Williams}, {Fender}, {Leung}, {Murphy}, \&
  {Green}}]{Sfaradi+2024}
{Sfaradi}, I., {Beniamini}, P., {Horesh}, A., {et~al.} 2024,
  \bibinfo{title}{{An off-axis relativistic jet seen in the long lasting
  delayed radio flare of the TDE AT 2018hyz},} \mnras, 527, 7672,
  \dodoi{10.1093/mnras/stad3717}

% type= article
\bibitem[{P. {Short} {et~al.}(2020){Short}, {Nicholl}, {Lawrence}, {Gomez},
  {Arcavi}, {Wevers}, {Leloudas}, {Schulze}, {Anderson}, {Berger}, {Blanchard},
  {Burke}, {Castro Segura}, {Charalampopoulos}, {Chornock}, {Galbany},
  {Gromadzki}, {Herzog}, {Hiramatsu}, {Horne}, {Hosseinzadeh}, {Howell},
  {Ihanec}, {Inserra}, {Kankare}, {Maguire}, {McCully}, {M{\"u}ller Bravo},
  {Onori}, {Sollerman}, \& {Young}}]{Short+2020}
{Short}, P., {Nicholl}, M., {Lawrence}, A., {et~al.} 2020, \bibinfo{title}{{The
  tidal disruption event AT 2018hyz - I. Double-peaked emission lines and a
  flat Balmer decrement},} \mnras, 498, 4119, \dodoi{10.1093/mnras/staa2065}

% type= article
\bibitem[{L. {Sironi} \& D. {Giannios}(2013){Sironi} \&
  {Giannios}}]{Sironi&Giannios2013}
{Sironi}, L., \& {Giannios}, D. 2013, \bibinfo{title}{{A Late-time Flattening
  of Light Curves in Gamma-Ray Burst Afterglows},} \apj, 778, 107,
  \dodoi{10.1088/0004-637X/778/2/107}

% type= article
\bibitem[{K. {Takahashi} {et~al.}(2022){Takahashi}, {Ioka}, {Ohira}, \& {van
  Eerten}}]{Takahashi+2022}
{Takahashi}, K., {Ioka}, K., {Ohira}, Y., \& {van Eerten}, H.~J. 2022,
  \bibinfo{title}{{Probing particle acceleration at trans-relativistic shocks
  with off-axis gamma-ray burst afterglows},} \mnras, 517, 5541,
  \dodoi{10.1093/mnras/stac3022}

% type= article
\bibitem[{G. {Taylor}(1950){Taylor}}]{Taylor1950}
{Taylor}, G. 1950, \bibinfo{title}{{The Formation of a Blast Wave by a Very
  Intense Explosion. I. Theoretical Discussion},} Proceedings of the Royal
  Society of London Series A, 201, 159, \dodoi{10.1098/rspa.1950.0049}

% type= article
\bibitem[{A. {Tchekhovskoy} {et~al.}(2014){Tchekhovskoy}, {Metzger},
  {Giannios}, \& {Kelley}}]{Tchekhovskoy+2014}
{Tchekhovskoy}, A., {Metzger}, B.~D., {Giannios}, D., \& {Kelley}, L.~Z. 2014,
  \bibinfo{title}{{Swift J1644+57 gone MAD: the case for dynamically important
  magnetic flux threading the black hole in a jetted tidal disruption event},}
  \mnras, 437, 2744, \dodoi{10.1093/mnras/stt2085}

% type= article
\bibitem[{O. {Teboul} \& B.~D. {Metzger}(2023){Teboul} \&
  {Metzger}}]{Teboul&Metzger2023}
{Teboul}, O., \& {Metzger}, B.~D. 2023, \bibinfo{title}{{A Unified Theory of
  Jetted Tidal Disruption Events: From Promptly Escaping Relativistic to
  Delayed Transrelativistic Jets},} \apjl, 957, L9,
  \dodoi{10.3847/2041-8213/ad0037}

% type= article
\bibitem[{S.~J. {Tingay} {et~al.}(1995){Tingay}, {Jauncey}, {Preston},
  {Reynolds}, {Meier}, {Murphy}, {Tzioumis}, {McKay}, {Kesteven}, {Lovell},
  {Campbell-Wilson}, {Ellingsen}, {Gough}, {Hunstead}, {Jonos}, {McCulloch},
  {Migenes}, {Quick}, {Sinclair}, \& {Smits}}]{Tingay+1995}
{Tingay}, S.~J., {Jauncey}, D.~L., {Preston}, R.~A., {et~al.} 1995,
  \bibinfo{title}{{Relativistic motion in a nearby bright X-ray source},} \nat,
  374, 141, \dodoi{10.1038/374141a0}

% type= article
\bibitem[{H. {van Eerten} {et~al.}(2010){van Eerten}, {Zhang}, \&
  {MacFadyen}}]{vanEerten+2010}
{van Eerten}, H., {Zhang}, W., \& {MacFadyen}, A. 2010,
  \bibinfo{title}{{Off-axis Gamma-ray Burst Afterglow Modeling Based on a
  Two-dimensional Axisymmetric Hydrodynamics Simulation},} \apj, 722, 235,
  \dodoi{10.1088/0004-637X/722/1/235}

% type= article
\bibitem[{S. {van Velzen} {et~al.}(2021){van Velzen}, {Gezari}, {Hammerstein},
  {Roth}, {Frederick}, {Ward}, {Hung}, {Cenko}, {Stein}, {Perley}, {Taggart},
  {Foley}, {Sollerman}, {Blagorodnova}, {Andreoni}, {Bellm}, {Brinnel}, {De},
  {Dekany}, {Feeney}, {Fremling}, {Giomi}, {Golkhou}, {Graham}, {Ho},
  {Kasliwal}, {Kilpatrick}, {Kulkarni}, {Kupfer}, {Laher}, {Mahabal}, {Masci},
  {Miller}, {Nordin}, {Riddle}, {Rusholme}, {van Santen}, {Sharma}, {Shupe}, \&
  {Soumagnac}}]{vanVelzen+2021}
{van Velzen}, S., {Gezari}, S., {Hammerstein}, E., {et~al.} 2021,
  \bibinfo{title}{{Seventeen Tidal Disruption Events from the First Half of ZTF
  Survey Observations: Entering a New Era of Population Studies},} \apj, 908,
  4, \dodoi{10.3847/1538-4357/abc258}

% type= article
\bibitem[{S.~C. {Wu} {et~al.}(2026){Wu}, {Tsuna}, {Mockler}, \&
  {Piro}}]{WuSamantha+2026}
{Wu}, S.~C., {Tsuna}, D., {Mockler}, B., \& {Piro}, A.~L. 2026,
  \bibinfo{title}{{Delayed Radio Emission in Tidal Disruption Events from
  Collisions of Outflows Driven by Disk Instabilities},} \apj, 998, 199,
  \dodoi{10.3847/1538-4357/ae36a3}

% type= article
\bibitem[{F. {Zhang} {et~al.}(2024){Zhang}, {Shu}, {Yang}, {Sun}, {Zhang},
  {Wang}, {Mou}, {Zhang}, {Zhou}, \& {Peng}}]{ZhangFabao+2024}
{Zhang}, F., {Shu}, X., {Yang}, L., {et~al.} 2024, \bibinfo{title}{{Delayed and
  Fast-rising Radio Flares from an Optical and X-Ray-detected Tidal Disruption
  Event in the Center of a Dwarf Galaxy},} \apjl, 962, L18,
  \dodoi{10.3847/2041-8213/ad1d61}

% type= article
\bibitem[{J. {Zhuang} {et~al.}(2025){Zhuang}, {Shen}, {Mou}, \&
  {Lu}}]{Zhuang+2025}
{Zhuang}, J., {Shen}, R.-F., {Mou}, G., \& {Lu}, W. 2025,
  \bibinfo{title}{{Interaction of an Outflow with Surrounding Gaseous Clouds as
  the Origin of Late-time Radio Flares in Tidal Disruption Events},} \apj, 979,
  109, \dodoi{10.3847/1538-4357/ad9b98}

% type= article
\bibitem[{J. {Zrake} {et~al.}(2018){Zrake}, {Xie}, \& {MacFadyen}}]{Zrake+2018}
{Zrake}, J., {Xie}, X., \& {MacFadyen}, A. 2018, \bibinfo{title}{{Radio Sky
  Maps of the GRB 170817A Afterglow from Simulations},} \apjl, 865, L2,
  \dodoi{10.3847/2041-8213/aaddf8}

\end{thebibliography}

\end{document}